\documentclass[10pt,twocolumn,letterpaper]{article}

\usepackage[pagenumbers]{cvpr} % To force page numbers, e.g. for an arXiv version

\usepackage[dvipsnames]{xcolor}

\usepackage{graphicx}
\usepackage{booktabs}
\usepackage{multirow}
\usepackage{amsmath}
\usepackage{algorithmic}
\usepackage{algorithm}
\usepackage{amsthm}
\usepackage{caption}

\usepackage{cuted}

\newtheorem{theorem}{Theorem}
\newtheorem{definition}{Definition}

\newcommand{\wm}{w}

\newcommand{\waterm}{^{w}}

\newcommand{\susp}{s}
\newcommand{\host}{o}
\newcommand{\model}{\boldsymbol{\epsilon}_{\boldsymbol{\theta}}}
\newcommand{\cone}{\boldsymbol{\phi}}
\newcommand{\ctwo}{\eta}
\newcommand{\mdp}{MDP}
\newcommand{\wdp}{WDP}

\definecolor{cvprblue}{rgb}{0.21,0.49,0.74}
\usepackage[pagebackref,breaklinks,colorlinks,citecolor=cvprblue]{hyperref}

\def\paperID{14369} % *** Enter the Paper ID here
\def\confName{CVPR}
\def\confYear{2024}

\title{Intellectual Property Protection of Diffusion Models via the Watermark Diffusion Process}
\author{
\vspace{2 mm}
Sen Peng$^{1}$\:
Yufei Chen$^{1}$\:\:
Cong Wang$^{1}$\:
Xiaohua Jia$^{1}$\\
$^{1}$City University of Hong Kong\\
}

\begin{document}
\maketitle
\begin{abstract}
\sloppy
Diffusion models have rapidly become a vital part of deep generative architectures, given today's increasing demands.
Obtaining large, high-performance diffusion models demands significant resources, highlighting their importance as intellectual property worth protecting.
However, existing watermarking techniques for ownership verification are insufficient when applied to diffusion models.
Very recent research in watermarking diffusion models either exposes watermarks during task generation, which harms the imperceptibility, or is developed for conditional diffusion models that require prompts to trigger the watermark.
This paper introduces WDM, a novel watermarking solution for diffusion models without imprinting the watermark during task generation.
It involves training a model to concurrently learn a Watermark Diffusion Process (WDP) for embedding watermarks alongside the standard diffusion process for task generation.
We provide a detailed theoretical analysis of WDP training and sampling, relating it to a shifted Gaussian diffusion process via the same reverse noise.
% Watermarks are extracted using a designated trigger, ensuring they stay unexposed during the primary task sampling.
% We further present a complete framework for verifying copyright infringement through hypothesis testing.
Extensive experiments are conducted to validate the effectiveness and robustness of our approach in various trigger and watermark data configurations.
The code can be found at the project page: \href{https://github.com/senp98/wdm}{https://github.com/senp98/wdm}.
\end{abstract}    
\section{Introduction}
\label{sec:intro}
\begin{figure*}[!ht]
\setlength{\abovecaptionskip}{0em}
\begin{center}
\includegraphics[width=\textwidth,trim=0 0 10 0,clip]{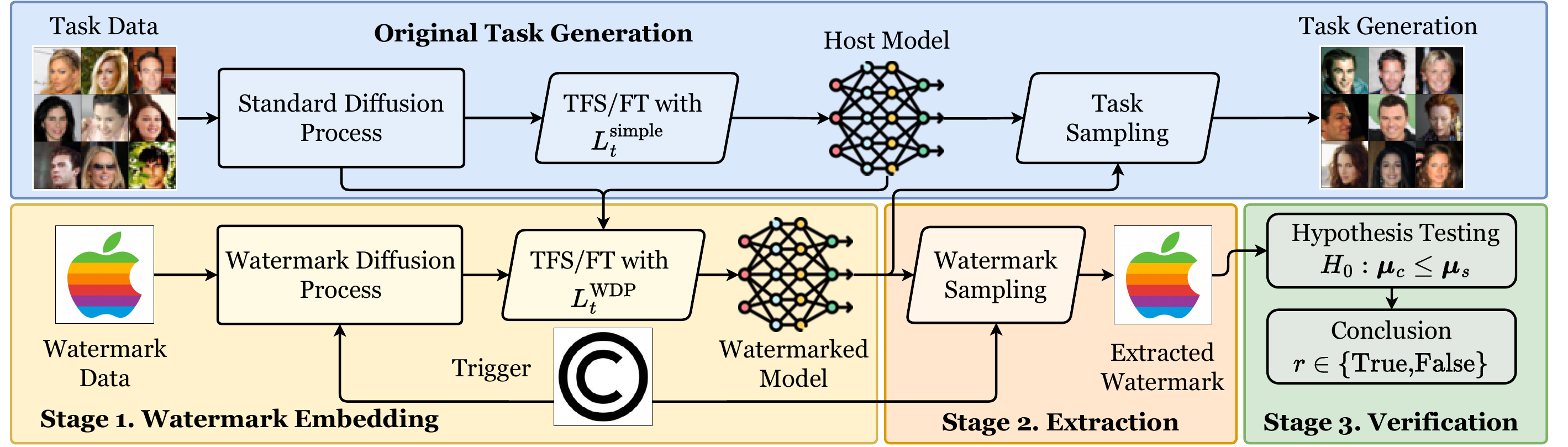}
\end{center}
\caption{The workflow of our proposed WDM. 
\textbf{Top}: The task data is learned by the standard diffusion process in the host model, which is trained or fine-tuned using the standard optimization objective in \cref{eq: l_simple}, and then sampled from the host model through the standard reverse process.
\textbf{Bottom}: The watermark data is initially embedded by training or fine-tuning the model to learn a WDP with the given trigger, using the optimization objective in \cref{eq: wdp_loss}. 
It can be extracted from the watermarked model using the same trigger and verified through similarity comparison and hypothesis testing, which leads to a conclusion regarding the model ownership.
}
\label{fig: framework}
\vspace{-1em}
\end{figure*}

Deep generative models are drawing significant attention in both academia and industry, driven by the growing demand for diverse applications.
Among them, diffusion models~\cite{sohl2015deep}, inspired by the diffusion process in non-equilibrium thermodynamics, have emerged as a notable architecture.
Recent studies have shown that diffusion models have outperformed other architectures such as the Generative Adversarial Network (GAN)~\cite{brock2018large,goodfellow2020generative} and Variational Auto-Encoder (VAE)~\cite{kingma2013auto,van2017neural} in terms of sampling quality and diversity~\cite{ho2020denoising,nichol2021improved,dhariwal2021diffusion}.
As various downstream services begin to incorporate well-trained large diffusion models like Stable Diffusion~\cite{rombach2022high} into their essential frameworks, the imperative to protect their Intellectual Property (IP) has intensified.
Firstly, these models represent valuable digital assets for developers, who invest significant time and financial resources in obtaining high-performance models~\cite{ramesh2021zero}.
Furthermore, the unauthorized dissemination of well-trained diffusion models can give rise to ethical concerns, such as the potential model misuse for generating disinformation.

Watermarking is a common strategy for model IP protection~\cite{li2021survey,peng2022intellectual}, which involves embedding specific information as the watermark into host models before deployment.
The embedded watermark in the suspected model can be extracted to confirm the ownership and verify whether there exists any IP violation.
% Static watermarking commonly utilizes regularization techniques to embed the specific information as the watermark into the model's static content, like its parameters or their distribution~\cite{uchida2017embedding,chen2019deepmarks,wang2021riga,liu2021watermarking}.
% In contrast, dynamic watermarking~\cite{le2020adversarial,yang2021robust,adi2018turning,zhang2018protecting,guo2018watermarking,li2019prove,darvish2019deepsigns,namba2019robust,jia2021entangled,szyller2021dawn} embeds a distinct pattern into the dynamic aspects of a model, such as its behavior (also called functionality), serving as the watermark.
% This approach typically initiates a backdoor attack to the host model first, and then utilizes the behavior pattern exhibited when triggering the implanted backdoor as the watermark.
While most existing studies focus on watermarking discriminative models, the development of watermarking techniques for generative models, especially diffusion models, still remains limited.
The challenge in watermarking generative models involves the complexity of manipulating the generation process without compromising the quality and diversity of the original task~\cite{ong2021protecting,zhao2023recipe}.
% and 2) the feature of unconditional generative models which operate without requiring any input.
%Previous studies have explored some watermarking techniques for GANs~\cite{fei2022supervised,ong2021protecting,yu2021artificial}
%However, these methods are not able to directly applied in diffusion models due to their different architectures.
Recent studies discussed the watermarking diffusion models ~\cite{zhao2023recipe,wdm2}.
However, these methods either are not applicable to unconditional diffusion models or focus on embedding watermarks into the task generation outcomes.

In this paper, we introduce WDM, a novel watermarking scheme designed for diffusion models through the Watermark Diffusion Process (WDP) without imprinting the watermark during task generation.
The idea of WDM lies in enabling the protected model to generate samples with a distinct data distribution different from the original task, serving as the watermark. 
The workflow of WDM mainly consists of three stages, as shown in cref{fig: framework}.
During the first stage, we embed the watermark data into the host model by concurrently learning a WDP by training or fine-tuning it, along with a standard diffusion process for the task data.
% During watermark embedding, we first sample a dataset from the target watermark distribution, which serves as the carrier for the watermark.
% We then develop the WDP, which is based on the standard diffusion process for samples in watermark data.
% The watermark data is embedded into the host model by training or fine-tuning it to concurrently learn two diffusion processes: the standard diffusion process for the task data and the WDP for the watermark data.
% We simplify the optimization objective for learning the WDP by demonstrating its equivalence to a diffusion process with the modified Gaussian kernel given specific configurations.
In the watermark extraction stage, we obtain the watermark data from the target model by utilizing the shared noise in the reverse diffusion process learned by WDP.
Lastly, the model ownership is verified through hypothesis testing, which involves the similarity comparison between the original and extracted watermark data.

To summarize, the primary contributions of this study are as follows:
1) We introduce a novel watermarking scheme for diffusion models via the watermark diffusion process, accompanied by its theoretical foundation and analysis that relates it to the standard diffusion process with the modified Gaussian kernel. 
2) We propose a complete framework for IP protection in diffusion models, which consists of the watermark embedding, extraction, and verification stages. 
3) We conduct extensive experiments to validate the effectiveness and robustness of our proposed method in defending against various types of attacks.

\section{Related Work}
\subsection{Diffusion Probabilistic Models}
Diffusion probabilistic models are generative models inspired by the diffusion process in non-equilibrium thermodynamics through a finite Markov chain~\cite{sohl2015deep}, which consists of the forward and reverse diffusion processes.
% The diffusion process characterizes data transition from one distribution to another through a finite Markov chain.
% The forward process diffuses the original data distribution into a noisy distribution through a forward transition kernel. 
% In contrast, the reverse process gradually removes noises from the noisy distribution to sample from the task data distribution. 
% % Given that each state transition in the forward diffusion process is sufficiently small, it can be proved that the transition kernels in the reverse process share the same form~\cite{sohl2015deep,song2020score}.
Subsequently, Ho \etal introduced Denoising Diffusion Probabilistic Models (DDPMs) with a simplified optimization objective for training diffusion models~\cite{ho2020denoising}.
Song et al. established a connection between the diffusion models and score-based models inspired by Langevin dynamics~\cite{song2019generative,song2020score}.
Further studies have focused on accelerating the sampling process~\cite{song2020denoising} and enhancing the data generalization capabilities~\cite{vahdat2021score}.

\subsection{Watermarking Discriminative Models}
Watermarking approaches for models mainly exploit the redundancy of the model's parameter space. 
% Parameters of the model are altered from one local optimal to another during watermark embedding, making it feasible to achieve without significantly affecting its original functionality.
Watermarking can be categorized into static and dynamic for discriminative models~\cite{peng2022intellectual}. 
Static watermarking~\cite{uchida2017embedding,chen2019deepmarks,wang2021riga,liu2021watermarking} defines the watermark as a specific pattern in its static content, such as a particular distribution of parameters. 
In contrast, dynamic watermarking~\cite{le2020adversarial,yang2021robust,adi2018turning,zhang2018protecting,guo2018watermarking,li2019prove,darvish2019deepsigns,namba2019robust,jia2021entangled,szyller2021dawn} designs the watermark as a specific pattern in model's dynamic contents, such as its behavior. 
It typically utilizes the backdoor attack~\cite{adi2018turning,zhang2018protecting,darvish2019deepsigns,szyller2021dawn,jia2021entangled} of models to construct the specific behavior as the watermark, where the watermark dataset contains trigger samples and target labels.

\subsection{Watermarking Generative Models}
Watermarking generative models is more challenging due to the difficulty of controlling the generative process. 
Previous studies~\cite{fei2022supervised,yu2021artificial,ong2021protecting} propose to watermark GANs by constructing mappings between trigger inputs and outputs given by the generator through regularization constraints. 
These methods are not applicable to diffusion models due to the different architectures.
Some studies are related to backdoor attacks in diffusion models~\cite{chou2022backdoor,trojdiff}.
They both introduce the concept of altering the transition kernel of the standard diffusion process to inject backdoors into diffusion models.
For comparison, our approach initially formulates the diffusion process with a modified Gaussian kernel (MDP) with a simplified optimization objective.
We then theoretically prove that a WDP is exactly a MDP under specific configurations. 
% Our approach focuses more on the watermarking application scenario, where we introduce a comprehensive framework for watermarking diffusion models.
Some recent studies~\cite{zhao2023recipe,wdm2} have also proposed watermarking methods for diffusion models, either by embedding the watermark into task generation results or extracting the watermark using specific prompts.
Notably, these methods are either unsuitable for unconditional diffusion models or expose the watermark during original task generation.
In contrast, our method generates the watermark data through a WDP given the corresponding trigger without affecting the standard diffusion process for task data generation. 
To the best of our knowledge, this is the first work that provides an ownership resolution of diffusion models by implanting a different diffusion process for the watermark data.

\section{Preliminary and Problem Definition}
\subsection{DDPMs}
In this paper, we choose the unconditional DDPM~\cite{ho2020denoising} as our target diffusion model since it simplifies the training objective in~\cite{sohl2015deep} and becomes a fundamental work for many subsequent studies.
The transition kernel in the forward diffusion process of DDPM has a Gaussian form:
\begin{equation}
\label{eq: sdp_kernel}
   q(\mathbf{x}_t|\mathbf{x}_{t-1})=\mathcal{N}(\mathbf{x}_t;\sqrt{\alpha_t}\mathbf{x}_{t-1},(1-\alpha_t)\mathbf{I}),
\end{equation}
where $\alpha_t=1-\beta_t$ and $\{\beta_t\in (0,1)\}_{t=1}^T$ is the variance schedule that controls the diffusion step size.
By denoting $\prod_{i=1}^{t}\alpha_i$ as $\Bar{\alpha}_t$, we can represent $\mathbf{x}_t$ using the reparameterization trick as 
\begin{equation}
\label{eq: sdp_xt}
    \mathbf{x}_t=\sqrt{\Bar{\alpha}_t}\mathbf{x}_0+\sqrt{1-\Bar{\alpha}_t}\boldsymbol{\epsilon},
\end{equation}
where $\boldsymbol{\epsilon} \sim \mathcal{N}(0,1)$.
% The reverse diffusion process also has a transition kernel with Gaussian form as 
% $q(\mathbf{x}_{t-1}|\mathbf{x}_{t},\mathbf{x}_{0})=\mathcal{N}(\mathbf{x}_{t-1};\boldsymbol{\Tilde{\mu}_{t}}(\mathbf{x}_t,\mathbf{x}_{0}),\Tilde{\sigma_t}\mathbf{I}).$
% For the reverse diffusion process, $p_{\theta}(\mathbf{x}_{t-1}|\mathbf{x}_t)=\mathcal{N}(\mathbf{x}_{t-1};\boldsymbol{\mu_{\theta}}(\mathbf{x}_t,t),\boldsymbol{\Sigma_{\theta}}(\mathbf{x}_t,t))$ is used to approximate this posterior probability.
% In DDPM, the variance of the approximated probability is fixed the same as $\boldsymbol{\Sigma_{\theta}}=\sigma_{t}=\beta_t\mathbf{I}$ for simplification. 
% Its mean can be predicted using a machine learning model $\model$ as 
% \begin{equation}
%      \boldsymbol{\mu_{\theta}}(\mathbf{x}_t,t)=\frac{1}{\sqrt{\alpha_t}}(\mathbf{x}_t-\frac{1-\alpha_t}{\sqrt{1-\Bar{\alpha}_t}}\model(\mathbf{x}_t,t)),
% \end{equation}
% where we can sample from the noisy distribution by 
% \begin{equation}
%     \mathbf{x}_{t-1}=\frac{1}{\sqrt{\alpha_t}}(\mathbf{x}_t-\frac{1-\alpha_t}{\sqrt{1-\Bar{\alpha}_t}}\model(\mathbf{x}_t,t))+\sigma_{t}\mathbf{z}
% \end{equation}
% with $\mathbf{z}\sim \mathcal{N}(0,1)$.
The training objective of $\model$ is similar to minimizing the negative log-likelihood in VAEs~\cite{kingma2013auto,van2017neural}, which can be optimized using the variational lower bound as
\begin{equation}
\label{eq: l_simple}
\scalebox{0.85}{
    $L_t^\text{simple}=\mathbb{E}_{t \sim [1, T], \mathbf{x}_0, \boldsymbol{\epsilon}_t} \Big[\|\boldsymbol{\epsilon}_t - \boldsymbol{\epsilon}_\theta(\sqrt{\bar{\alpha}_t}\mathbf{x}_0 + \sqrt{1 - \bar{\alpha}_t}\boldsymbol{\epsilon}_t, t)\|^2 \Big]$},
\end{equation}
where $\boldsymbol{\epsilon}_t$ is the reverse noise at $t$.

\subsection{Threat Model}
We investigate the threat model of detecting and verifying the diffusion model stealing in this paper.
Specifically, we consider the scenario in which an owner trains a diffusion model on task data and obtains a host model $\model^{\host}$ for deployment.
An attacker who has successfully stolen a copy of the deployed model proceeds to claim its copyright without acknowledging $\model^{\host}$ from the model owner.
To detect and identify the possible IP violation of a suspected model, the owner instead trains or fine-tunes $\model^{\host}$ to obtain the watermarked model, denoted as $\model^{\wm}$ for deployment.% using a watermark dataset $\mathcal{D}_{\text{wm}}$ .
% The owner keeps both the embedded watermark and the trigger used as secrets.
Once obtained a suspected model $\model^{\susp}$ (stolen from $\model^{\wm}$) from a potential attacker, the owner retrieves the watermark and verify if it is stolen by comparing the extracted watermark against the initially embedded one.
In this context, we assume that the owner can gain complete access to the suspected model for watermark detection and verification.
Alternatively, this whole process can be delegated to a trusted third-party.
The choice of the white-box setting is due to the fact that unconditional diffusion models require no input for generation. 
This makes it impractical to extract an embedded watermark without accessing the suspected model, unless the watermark is integrated with the original task generation results which harms the watermark imperceptibility.

\subsection{Defense Goals} 
We aim to deign a watermarking method for diffusion models given dataset $\mathcal{D}_{\text{train}}$ and $\mathcal{D}_{\text{wm}}$ to enables the owner for detecting the model stealing from attackers.
We also seek to develop a complete watermarking framework $\mathcal{F}_{\wm}$ which consists of $\{f_{\text{embed}},f_{\text{extract}},f_{\text{verify}}\}$.
Specifically, it includes the following three sub-algorithms:
% \textbf{1) Watermark Embedding} $f_{\text{embed}}$: It embeds the watermark data $\mathbf{a}$ into the host model $\model^{\host}$ to obtain the watermarked model $\model^{\wm}$ with the trigger $\mathbf{b}$ as $f_{\text{embed}}(\model^{\host},\mathbf{a},\mathbf{b})\rightarrow \model^{\wm}.$
% \textbf{2) Watermark Extraction} $f_{\text{extract}}$: It extracts the watermark data $\hat{\mathbf{a}}$ from the suspected model using the trigger $\mathbf{b}$ as $f_{\text{extract}}(\model^{\susp},\mathbf{b})\rightarrow \hat{\mathbf{a}}$ for a suspected model $\model^{\susp}$.
% \textbf{3) Watermark Verification} $f_{\text{verify}}$: It compares the extracted watermark data $\hat{\mathbf{a}}$ with the original $\mathbf{a}$ and gives a verification result $r$ as $f_{\text{verify}}(\mathbf{a},\hat{\mathbf{a}})\rightarrow r \in \{\text{True},\text{False}\}.$
\begin{enumerate}[leftmargin=*,topsep=0pt]
    \setlength{\itemsep}{1pt}
    \item \textbf{Watermark Embedding} $f_{\text{embed}}$: It embeds the watermark data $\mathbf{a}$ into the host model $\model^{\host}$ to obtain the watermarked model $\model^{\wm}$ with the trigger $\mathbf{b}$ as $f_{\text{embed}}(\model^{\host},\mathbf{a},\mathbf{b})\rightarrow \model^{\wm}.$
    \item \textbf{Watermark Extraction} $f_{\text{extract}}$: It extracts the watermark data $\hat{\mathbf{a}}$ from the suspected model using the trigger $\mathbf{b}$ as $f_{\text{extract}}(\model^{\susp},\mathbf{b})\rightarrow \hat{\mathbf{a}}$ for a suspected model $\model^{\susp}$.
    \item \textbf{Watermark Verification} $f_{\text{verify}}$: It compares the extracted watermark data $\hat{\mathbf{a}}$ with the original $\mathbf{a}$ and gives a verification result $r$ as $f_{\text{verify}}(\mathbf{a},\hat{\mathbf{a}})\rightarrow r \in \{\text{True},\text{False}\}.$
\end{enumerate}
In addition to the aforementioned expectations, the proposed watermarking scheme also need to satisfy the defense goals as follows: 
% \textbf{1) Model Fidelity}: The watermarking method should not significantly affect the performance of diffusion models in their original task generation.
% \textbf{2) Watermark Fidelity}: The watermark can be extracted from the watermarked model it embeds, and its existence can be successfully verified.
% \textbf{3) Watermark Robustness}: The embedded watermark is resistant to watermark removing attacks. In this study, we consider the removing attacks as model transformation attacks and model fine-tuning attacks.
\begin{enumerate}[leftmargin=*,topsep=0pt]
    \setlength{\itemsep}{1pt}
    \item \textbf{Model Fidelity}: The watermarking method should not significantly affect the performance of diffusion models in their original task generation.
    \item \textbf{Watermark Fidelity}: The watermark can be extracted from the watermarked model it embeds, and its existence can be successfully verified.
    \item \textbf{Watermark Robustness}: The embedded watermark is resistant to watermark removing attacks. In this study, we consider the removing attacks as model transformation attacks and model fine-tuning attacks.
\end{enumerate}
The development of such a complete watermarking technique can provide the necessary security for diffusion models, enabling the safe distribution to third parties without IP violations.
\begin{figure*}[htbp]
\setlength{\abovecaptionskip}{0em}
\begin{center}
\includegraphics[width=\textwidth,trim=0 5 30 2,clip]{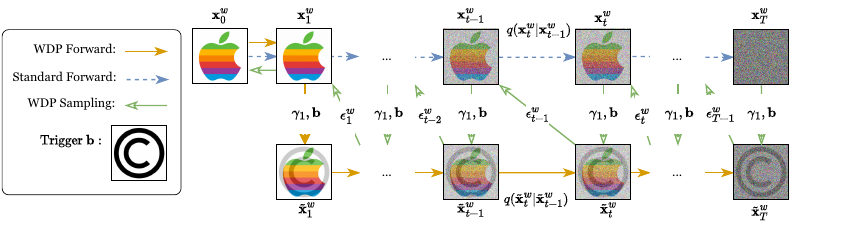}
\end{center}
\caption{Illustration of the watermark diffusion process (WDP). 
The standard forward diffusion process diffuses the data $\mathbf{x}\waterm_{0}$ (the apple icon) to a noisy distribution.
The forward WDP for $\mathbf{x}\waterm_{t}$ is established based on its standard forward diffusion process, as shown in \cref{eq: wdp}, using the provided trigger (the copyright icon).
Sampling for the WDP is accomplished by employing the standard reverse diffusion process with the same noise $\boldsymbol{\epsilon}_t$ obtained from its WDP.}
\label{fig: wdp}
\vspace{-1em}
\end{figure*}

\section{Proposed Method}
This section introduces our proposed Watermarking Diffusion Models (WDM) method. 
% The particular behavior of a diffusion model that can generate watermark data follows the distribution $q(\mathbf{x}\waterm)$ is used as the watermark to embed.
In practice, we use the watermark dataset $\mathcal{D}_{\text{wm}}$ consisting of samples $\mathbf{x}\waterm_0 \sim q(\mathbf{x}\waterm)$ from the distribution as the carrier.
First, we illustrate the diffusion process with a modified Gaussian kernel, denoted as MDP, and demonstrate its simplified optimization objective theoretically.
We then define the WDP and prove that a WDP is an MDP under specific configurations. 
The watermark data is embedded into the host model by training or fine-tuning with the designed optimization objective for concurrently learning the WDP and the standard diffusion process. 
% This objective enables the model to concurrently learn a standard diffusion process for task data and a WDP for watermark data.
% The trigger for embedded watermark data can either be chosen based on prior knowledge or randomly generated through a pseudo-random function with the secret key.
Since the WDP shares the same reverse noise as the standard diffusion process it is constructed upon, watermark extraction can be considered as the standard reverse process using the shared noise predicted by the learned WDP.
Lastly, the extracted watermark is verified through similarity comparison and hypothesis testing.
\cref{fig: wdp} demonstrates the relationship between a WDP and the standard diffusion process it is based upon.

\subsection{Modified Diffusion Process}
Inspired by the idea of multiplying distributions discussed in~\cite{sohl2015deep}, we first define the diffusion process with the modified Gaussian kernel (modified diffusion process) for data distribution $q(\mathbf{x})$.
\begin{definition}
\label{definition: mdp}
For samples $\mathbf{x}_0 \sim q(\mathbf{x})$, we define the diffusion process in finite timesteps $\{1,...,T\}$, whose forward transition kernel can be represented as
\begin{equation}
\label{eq: mdp}
     q(\mathbf{x}_t|\mathbf{x}_{t-1})=\mathcal{N}(\mathbf{x}_t;\sqrt{\alpha_t}(\mathbf{x}_{t-1}+\cone_t),\ctwo^{2}(1-\alpha_t)\mathbf{I}),
\end{equation}
as a diffusion process with a modified Gaussian kernel. $\cone_t$ is the constant schedule and $\ctwo\in[0,1]$.
\end{definition}
\noindent To get the reverse transition kernel $q(\mathbf{x}_{t-1}|\mathbf{x}_{t})$ of an MDP, we use $p_{\theta}(\mathbf{x}_{t-1}|\mathbf{x}_{t})$ to approximate this probability, learned by a model $\model$, to predict the reverse noise $\boldsymbol{\epsilon}_t$ in the reverse diffusion process.
This handling follows the same technique in~\cite{ho2020denoising}.
The simplified optimization objective for training $\model$ to learn an MDP can be demonstrated as follows:
\begin{theorem}
\label{theorem: mdp}
For the model $\model$ used to predict the noise $\boldsymbol{\epsilon}_t$ in the reverse diffusion process of an MDP, the optimization objective at timestep $t$ for its training can be simplified as the same form in \cref{eq: l_simple}
\begin{equation}
\label{eq: mdp_loss}
\scalebox{0.85}{$L^{\text{\mdp}}_t=\mathbb{E}_{t \sim [1, T], \mathbf{x}_0, \boldsymbol{\epsilon}_t} \Big[\|\boldsymbol{\epsilon}_t - \model(\sqrt{\bar{\alpha}_t}\mathbf{x}_0 + \sqrt{1 - \bar{\alpha}_t}\boldsymbol{\epsilon}_t, t)\|^2 \Big]$}.
\end{equation}
\end{theorem}
\noindent The proof of \cref{theorem: mdp} is demonstrated in Sec.7 in Appendix.

\subsection{Watermark Diffusion Process}
We construct the watermark diffusion process based on the forward transition kernel of the standard one, demonstrated in \cref{eq: sdp_kernel}, for samples $\mathbf{x}_{0} \sim q(\mathbf{x})$.
\begin{definition}
\label{definition: wdp}
For samples $\mathbf{x}_0 \sim q(\mathbf{x})$, let $\mathbf{x}_{1:T}$ denote the states within a standard diffusion process with the forward transition kernel specified as \cref{eq: sdp_kernel}.
We define the diffusion process, whose state $\Tilde{\mathbf{x}}_{t}$ for $t\in\{0,...,T\}$ satisfies:
\begin{equation}
\label{eq: wdp}
     \Tilde{\mathbf{x}}_{t}=\gamma_1{\mathbf{x}_{t}}+(1-\gamma_1)\mathbf{b},
\end{equation}
as a watermark diffusion process, where the constant $\mathbf{b}\in \mathbb{R}^{\|\mathbf{x}\|}$ denotes the trigger and $\gamma_1\in[0,1]$ denotes the trigger factor.
\end{definition}
\noindent The trigger $\mathbf{b}$ drifts the mean of state distribution in the diffusion process.
The state in $\Tilde{\mathbf{x}}_{0:T}$ of the WDP shares the same reverse noise $\boldsymbol{\epsilon}$ with that in the standard diffusion process upon which it is based, since we can get the following relation from \cref{eq: wdp}:
% \begin{align}
%     \Tilde{\mathbf{x}}_{t}
%     &=\gamma_1\mathbf{x}_{t}+(1-\gamma_1)\mathbf{b}\\
%     &=\gamma_1\sqrt{\Bar{\alpha}_t}\mathbf{x}_{0}+\gamma_1\sqrt{1-\Bar{\alpha}_t}\boldsymbol{\epsilon}+(1-\gamma_1)\mathbf{b}.
% \end{align}
\begin{equation}
    \Tilde{\mathbf{x}}_{t}=\gamma_1\sqrt{\Bar{\alpha}_t}\mathbf{x}_{0}+\gamma_1\sqrt{1-\Bar{\alpha}_t}\boldsymbol{\epsilon}+(1-\gamma_1)\mathbf{b}.
\end{equation}
The shared reverse noise $\boldsymbol{\epsilon}_t$ indicates that sampling from the data distribution learned by the WDP can be accomplished via the standard reverse diffusion process upon which it is based. 
% Thus, by making a diffusion model to concurrently learn a WDP for the watermark data and a standard diffusion process for the task data, we can only sample the watermark data when giving the reverse noise predicted by the WDP.
Thus, we can only sample the watermark data when giving the reverse noise predicted by the WDP.
This is achieved without affecting the task generation results when giving the reverse noise predicted by the standard diffusion process.
To establish the optimization objective for learning a WDP, we connect the WDP with the MDP under specific configurations, detailed as follows
\begin{theorem}
\label{theorem: wdp}
For samples $\mathbf{x}_0 \sim q(\mathbf{x})$, the WDP defined by \cref{eq: wdp} is exactly an MDP in \cref{eq: mdp} with the following configuration satisfied: 
\begin{equation}
\label{eq: cwt}
   \cone_t=(\frac{1}{\sqrt{\alpha_t}}-1)(1-\gamma_1)\mathbf{b} \qquad\text{and}\qquad \ctwo=\gamma_1.
\end{equation}
\end{theorem}
\noindent The proof of \cref{theorem: wdp} can be found in Sec.7 in Appendix. 
It indicates that we can train the model $\model$ to learn a WDP using the objective in \cref{theorem: mdp} derived from the MDP under the specific configuration. 
With this, we have established the complete theoretical framework and analysis of our proposed WDM.

\subsection{Watermark Embedding}
We embed the watermark by training the diffusion model to concurrently learn the task data through a standard diffusion process and the watermark data through a WDP.
% The significance of this design is that task sampling does not generate samples in the watermark data distribution, which prevents the watermark from leaking and preserves the model's functionality.
Details of our watermark embedding algorithm are demonstrated in \cref{alg: embedding}. 
Watermark embedding is achieved by training from scratch or fine-tuning the host model $\model^{o}$.
During each iteration, we first sample $\mathbf{x}_0 \in \mathcal{D}_{\text{train}}$ and $\mathbf{x}\waterm_0 \in \mathcal{D}_{\text{wm}}$.
We then sample the noises $\boldsymbol{\epsilon}$ and $\boldsymbol{\epsilon}\waterm$ in this diffusion process separately from $\mathcal{N}(\mathbf{0},\mathbf{I})$ and the timestep $t \in \text{Uniform}(\{1,...,T\})$.
For the task sample $\mathbf{x}_0$, we compute its corresponding state $\mathbf{x}_t$ at timestep $t$ within its standard diffusion process.
For the watermark sample $\mathbf{x}\waterm_0$, we initially compute its state $\mathbf{x}\waterm_t$ in the standard diffusion process. 
Subsequently, the corresponding state $\Tilde{\mathbf{x}}\waterm_t$ in its WDP is constructed based on $\mathbf{x}\waterm_t$ as:
\begin{equation}
    \Tilde{\mathbf{x}}\waterm_t=\gamma_{1}(\sqrt{\bar{\alpha}_t}\mathbf{x}\waterm_0+\sqrt{1-\bar{\alpha}_t}\boldsymbol{\epsilon}\waterm)+(1-\gamma_{1})\mathbf{b}.
\end{equation}
Here $\gamma_{1}$ denotes the trade-off factor for balancing the impact of the trigger $\mathbf{b}$.
According to \cref{theorem: mdp} and \cref{theorem: wdp}, we can obtain the training optimization objective for jointly learning the standard process and WDP, which is also the loss function for watermark embedding:
\begin{equation}
\label{eq: wdp_loss}
\scalebox{0.78}{$L^{\text{\wdp}}_{t}\!=\!\mathbb{E}_{t \sim [1, T], \mathbf{x}_0, \mathbf{x}^{\waterm}_0, \boldsymbol{\epsilon}_t} \Big[\gamma_{2}\|\boldsymbol{\epsilon}\!-\!\boldsymbol{\epsilon_{\theta}}(\mathbf{x}_{t},t)\|^2\!+\!\|\boldsymbol{\epsilon}\waterm\!-\!\boldsymbol{\epsilon_{\theta}}(\Tilde{\mathbf{x}}\waterm_t,t)\|^2\Big]$},
\end{equation}
where $\gamma_{2}$ is a trade-off factor for balancing the standard and watermark diffusion processes.
\begin{algorithm}[htbp]
   \caption{Watermark Embedding}
   \label{alg: embedding}
\begin{algorithmic}[1]
   \STATE {\bfseries Input:} model $\model$, training dataset $\mathcal{D}_{\text{train}}$, watermark dataset $\mathcal{D}_{\text{wm}}$, trigger $\mathbf{b}$, factor $\gamma_{1}$, $\gamma_{2}$ and diffusion step $T$.
   \REPEAT
   \STATE $\mathbf{x}_{0}\sim \mathcal{D}_{\text{train}}$
   \STATE $\mathbf{x}\waterm_{0}\sim \mathcal{D}_{\text{wm}}$
   \STATE $t \sim \text{Uniform}(\{1,...,T\})$
   \STATE $\boldsymbol{\epsilon},\boldsymbol{\epsilon}\waterm\sim \mathcal{N}(\mathbf{0},\mathbf{I})$
   \STATE $\mathbf{x}_{t}=\sqrt{\bar{\alpha}_t}\mathbf{x}_0+\sqrt{1-\bar{\alpha}_t}\boldsymbol{\epsilon}$
   \STATE $\mathbf{x}\waterm_{t}=\sqrt{\bar{\alpha}_t}\mathbf{x}\waterm_0+\sqrt{1-\bar{\alpha}_t}\boldsymbol{\epsilon}\waterm$
   %\STATE $\mathbf{x}\modi_{t}=$
   \STATE Take gradient descent step on: \\ \scalebox{0.8}{$\nabla_{\boldsymbol{\theta}}(\gamma_{2}\|\boldsymbol{\epsilon}-\model(\mathbf{x}_{t},t)\|^2+\|\boldsymbol{\epsilon}\waterm-\model(\gamma_{1}\mathbf{x}\waterm_{t}+(1-\gamma_{1})\mathbf{b},t)\|^2)$}
   \UNTIL{converged}
\end{algorithmic}
\end{algorithm}

\begin{algorithm}[htbp]
   \caption{Watermark Extraction}
   \label{alg: extraction}
\begin{algorithmic}[1]
   \STATE {\bfseries Input:} model $\model$, trigger $\mathbf{b}$, trigger factor $\gamma_{1}$, diffusion step $T$.
   \STATE $\mathbf{x}^{\wm}_{T} \sim \mathcal{N}(0,1)$
   \FOR{$t=T,...,1$}
   \STATE $\mathbf{z}\sim \mathcal{N}(\mathbf{0},\mathbf{I})$ {\bfseries if} $t>1$ {\bfseries else} $\mathbf{z}=0$
   \STATE \scalebox{0.85}{$\mathbf{x}\waterm_{t-1}=\frac{1}{\sqrt{\alpha_t}}(\mathbf{x}\waterm_t-\frac{1-\alpha_t}{\sqrt{1-\Bar{\alpha}_t}}\model(\gamma_{1}\mathbf{x}\waterm_t+(1-\gamma_{1})\mathbf{b},t))+\sigma_{t}\mathbf{z}$}
   \ENDFOR
\end{algorithmic}
\end{algorithm}

\subsection{Watermark Extraction}
Since the standard diffusion process and WDP upon which it is based share the same reverse noise, watermark extraction can be achieved through the standard reverse diffusion process using the noise learned in the WDP.
The detailed algorithm for watermark extraction is shown in \cref{alg: extraction}.
For a model $\model$, given the trigger $\mathbf{b}$ and trigger factor $\gamma_{1}$, we first sample $\mathbf{x}^{\wm}_{T}$ from $\mathcal{N}(0,1)$.
We then compute its corresponding state $\Tilde{\mathbf{x}}\waterm_{t}$ in the WDP as $\Tilde{\mathbf{x}}\waterm_{t} = \gamma_{1}\mathbf{x}\waterm_t+(1-\gamma_{1})\mathbf{b}$.
Finally, $\Tilde{\mathbf{x}}\waterm_{t}$ is used as an input to the model $\model$ for obtaining the shared reverse noise and the $\mathbf{x}\waterm_{t-1}$. 
The trigger $\mathbf{b}$ and trigger factor $\gamma_1$ should be selected so there is enough divergence between the state distribution in the WDP and distribution in the standard one. 
We discuss the selection of suitable triggers and factors in Sec.8 in Appendix.

\subsection{Watermark Verification}
To verify the existence of the embedded watermark, we develop a two-step verification procedure, including watermark similarity comparison and hypothesis testing.

\noindent \textbf{Similarity Comparison}
We first measure the similarity between the extracted watermark from the suspected model and the original embedded one. 
When the watermark dataset contains only one sample, which means the variance of the watermark data equals 0, we employ the Structural Similarity Index (SSIM) between the embedded watermark sample and an extracted one from the suspected model as their similarity. 
In cases where the watermark dataset contains multiple samples, we use the FID score between the sampled batch from the watermark model and that from the watermark data as their similarity. 

\noindent \textbf{Hypothesis Testing}
Hypothesis testing is used in our proposed WDM to conclude whether a suspected model is stolen from the host one.
It is essential to obtain the watermark similarity $\boldsymbol{\mu_s}$ from above and compare it with the similarity value $\boldsymbol{\mu_c}$ from a reference batch. 
The reference batch is introduced in this context by adding the Gaussian noise $\mathcal{N}(0,\sigma_r)$ to the embedded watermark data. 
In the case of the single watermark dataset setting, we set $\sigma_r=0.2$.
We also set $\sigma_r=0.05$ for the multiple-sample setting.
The smaller similarity value indicates that the difference between the extracted watermark data and the original one is smaller.
The null hypothesis is formulated as $H_0:\boldsymbol{\mu}_{c} \geq \boldsymbol{\mu}_{s}$ for using SSIM as similarity (since larger SSIM represents higher similarity), implying that the suspected model is stolen from the host model. 
The alternative hypothesis is established as $H_1:\boldsymbol{\mu}_{c} < \boldsymbol{\mu}_{s}$.
The direction of hypothesis reverses when taking the FID as the similarity since smaller FID.
We take the significance level of $\alpha=0.01$ in our analysis to minimize the risk of a Type-I error.
%================================================= 
\section{Experiments}
%================================================= 
\subsection{Experimental Settings}
Our experiments utilize two benchmark datasets for task generation: CIFAR-10~\citep{krizhevsky2009learning} and CelebA~\citep{liu2018large}. 
The implementation of our WDM follows the diffusion model architecture described in~\cite{nichol2021improved}.
The baseline models are obtained by training for 300K iterations on each task dataset.
Subsequently, the watermark data is embedded into the baseline models through fine-tuning.
To evaluate the model fidelity, we generate the batch with 50k samples to compare against the reference batch, using the IS~\citep{isscore} and FID~\citep{heusel2017gans} as the metrics.
Additionally, we generate 100 batches, each containing 100 watermark samples, for sWDM setting and ten batches with 10k watermark samples for mWDM setting.
All training, fine-tuning, and sampling procedures are conducted using 8 NVIDIA RTX A5000 GPUs.

We evaluate six WDM settings, including three types of triggers and two watermark datasets as shown in \cref{tab: exp_setting}.
The trigger used in watermark embedding can be either specifically selected (using the copyright icon) or randomly generated through a pseudo-random function given the secret key (using the pixel-level and column-level random white and black images).
The watermark data distribution can have a zero variance, indicating that the watermark data contains only one single sample (using the apple icon).
Alternatively, it can also have a non-zero variance, which corresponds that the watermark data contains multiple samples (using the samples of class "5" in MNIST training dataset~\cite{lecun1995learning}).
Detailed information regarding our WDM settings and implementations are demonstrated in Sec.9 in Appendix.
\begin{table}[htbp]
\centering
\setlength{\tabcolsep}{3pt}
\setlength{\fboxsep}{0pt}
\scalebox{1}{
\begin{tabular}{@{}lccc@{}}
\toprule
Settings & sWDM-sel & sWDM-randp & sWDM-randc \\ \midrule
Trigger & \raisebox{-.5\height}{\fbox{\includegraphics[width=25pt]{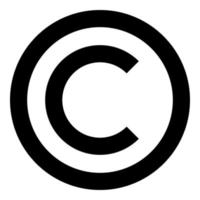}}} & \raisebox{-.5\height}{\fbox{\includegraphics[width=25pt]{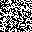}}} & \raisebox{-.5\height}{\fbox{\includegraphics[width=25pt]{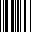}}} \\ \midrule
\begin{tabular}[c]{@{}l@{}}Watermark\\ Dataset\end{tabular} & \raisebox{-.5\height}{\fbox{\includegraphics[width=25pt]{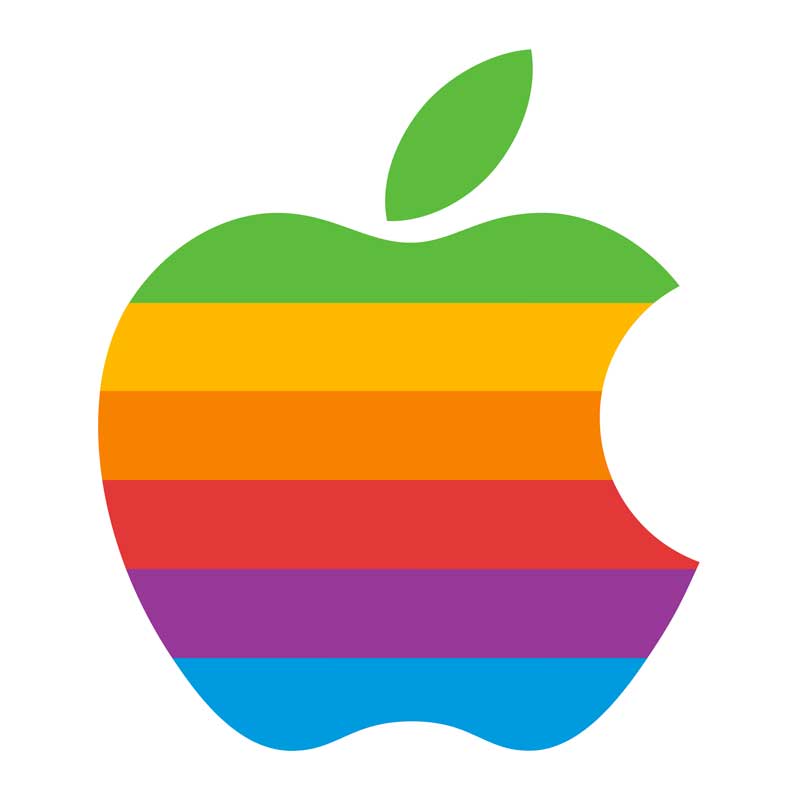}}} & \raisebox{-.5\height}{\fbox{\includegraphics[width=25pt]{figures/single_wm.jpg}}} & \raisebox{-.5\height}{\fbox{\includegraphics[width=25pt]{figures/single_wm.jpg}}} \\ \midrule \midrule
Settings & mWDM-sel & mWDM-randp & mWDM-randc \\ \midrule
Trigger & \raisebox{-.5\height}{\fbox{\includegraphics[width=25pt]{figures/trigger_sel.jpg}}} & \raisebox{-.5\height}{\fbox{\includegraphics[width=25pt]{figures/trigger_randp.png}}} & \raisebox{-.5\height}{\fbox{\includegraphics[width=25pt]{figures/trigger_randc.png}}} \\ \midrule
\begin{tabular}[c]{@{}l@{}}Watermark\\ Dataset\end{tabular} & \raisebox{-.5\height}{\fbox{\includegraphics[width=25pt]{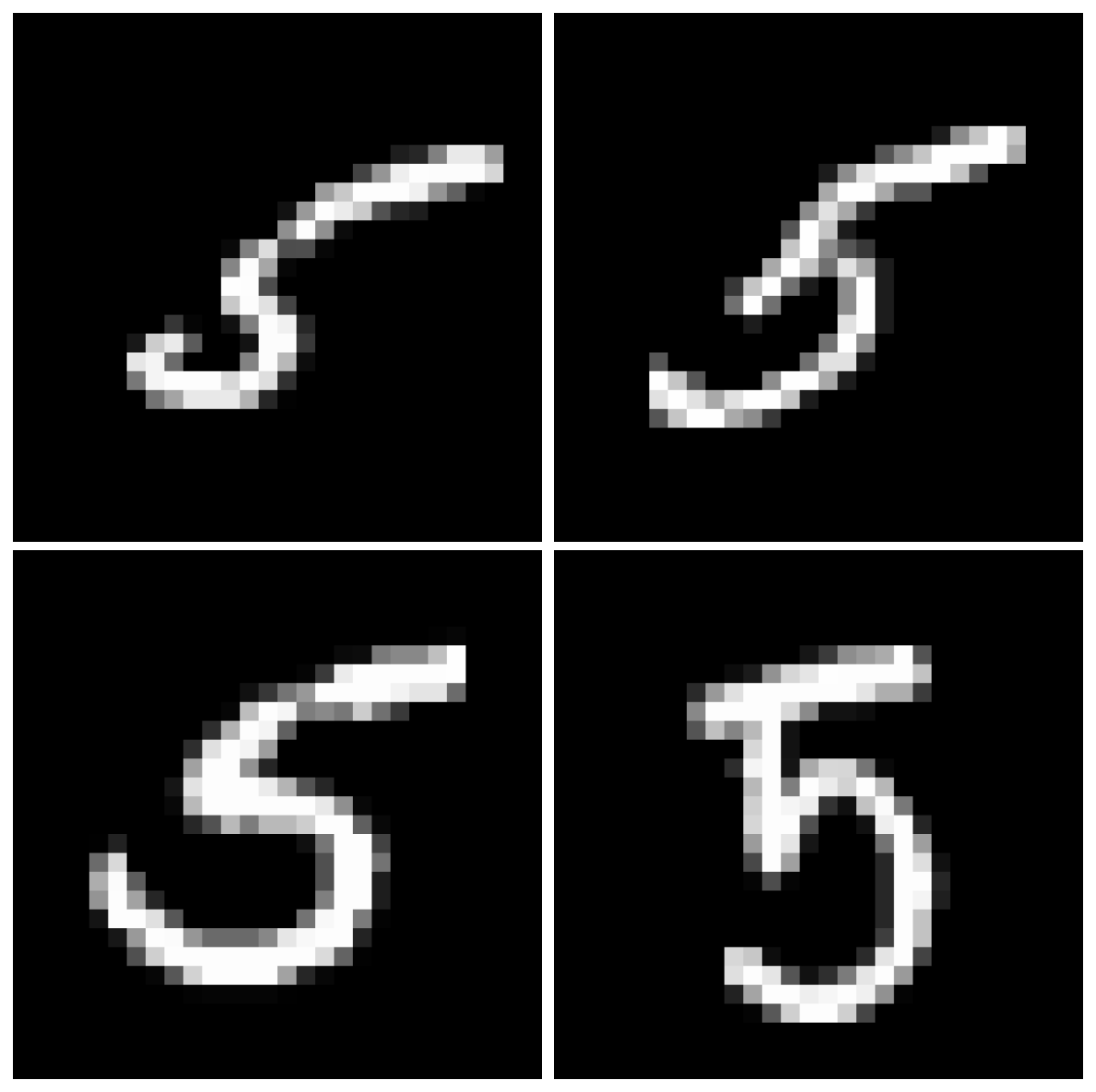}}} & \raisebox{-.5\height}{\fbox{\includegraphics[width=25pt]{figures/multiple_wm.pdf}}} & \raisebox{-.5\height}{\fbox{\includegraphics[width=25pt]{figures/multiple_wm.pdf}}} \\ \bottomrule
\end{tabular}
}
\caption{Demonstration of different WDM settings in our experiments. We evaluate six WDM settings, including three types of triggers and two types of watermark data. 
We respectively denote the WDM with the single-sample watermark dataset (the apple icon) and multiple-sample one (class "5" in the MNIST training dataset) as sWDM and mWDM. 
We also denote the selected trigger as WDM-sel (the copyright icon), the randomly generated trigger at pixel level as WDM-randp, and randomly generated trigger at the column level as WDM-randc.}
\label{tab: exp_setting}
\vspace{-1em}
\end{table}

\subsection{Model and Watermark Fidelity}
\begin{table}
\setlength{\tabcolsep}{2pt}
\centering
\scalebox{0.78}{
\begin{tabular}{@{}l|cccc|cccc@{}}
\toprule
 & \multicolumn{4}{c|}{CIFAR-10} & \multicolumn{4}{c}{CelebA} \\ \midrule
Methods & IS $\uparrow$ & FID$\downarrow$ & WS & VC$\downarrow$ & IS $\uparrow$ & FID$\downarrow$ & WS & VC$\downarrow$ \\ \midrule
Baseline & \bf{9.27} & \bf{5.13} & NA & NA & \bf{2.66} & \bf{1.72} & NA & NA \\ \midrule
% Baseline-ft & 9.02 & 5.49 & 0.029/358 & 1.00/1.00 & 2.62 & 1.94 & 0.001/371 & 1.00/1.00 \\ \midrule
%sWDM-ref & - & - & 0.000 & 0.000 & - & - & 0.000 & 0.000 \\
Baseline-ft & \bf{9.02} & \bf{5.49} & 0.029 & 1.00 & 2.62 & 1.94 & 0.001 & 1.00 \\ 
sWDM-sel & 9.01& 5.71 & 0.997 & $<\!10^{-324}$ & 2.63 & 2.00 & 0.998 & $<\!10^{-324}$ \\
sWDM-randp & 8.94 & 5.68 & 0.997 & $<\!10^{-324}$ & 2.60 & 2.05 & 0.997 & $<\!10^{-324}$ \\
sWDM-randc & 9.04 & 5.84 & \bf{0.998} & $\bf{<\!10^{-324}}$ & 2.61 & 2.11 & \bf{0.998} & $\bf{<\!10^{-324}}$ \\
sWDM-avg & 8.99 & 5.74 & 0.998 & $<\!10^{-324}$ & 2.61 & 2.05 & 0.998 & $<\!10^{-324}$ \\ \midrule
%mWDM-ref & - & - & 0.000 & 0.000 & - & - & 0.000 & 0.000 \\
Baseline-ft & \bf{9.02} & \bf{5.49} & 358 & 1.00 & 2.62 & 1.94 & 371 & 1.00 \\ 
mWDM-sel & 8.98 & 6.03 & 6.78 & $10^{-28}$ & 2.62 & 1.98 & 8.80  & $10^{-28}$ \\
mWDM-randp & 8.91 & 6.38 & 7.41 &  $10^{-27}$  & 2.61 & 2.05 & 7.92 & $10^{-28}$ \\
mWDM-randc & 9.01 & 5.89 & \bf{5.91} & $\bf{10^{-28}}$ & 2.59 & 2.11 & \bf{5.64} & $\bf{10^{-28}}$ \\
mWDM-avg & 8.97 & 6.10 & 6.70 & $10^{-28}$ & 2.61 & 2.05 & 7.45 & $10^{-28}$ \\ \bottomrule
\end{tabular}
}
\caption{Model and watermark fidelity in different WDM settings. 
The sWDM-avg in this table represents the average value of sWDM-sel, sWDM-randp, and sWDM-randc, with a similar interpretation for mWDM-avg.
Given that the smallest numerical value in our experimental device is at the level $10^{-324}$, we denote the numeric results, which equals 0 with $<\!10^{-324}$.
The best values are highlighted in bold.}
\label{tab: model_fidelity}
\vspace{-1em}
\end{table}
\cref{tab: model_fidelity} presents both model and watermark fidelity in various WDM settings compared to the baseline and fine-tuned baseline host models. 
The results indicate that the model fidelity degradation resulting from watermark embedding across all WDM settings is very small.
On average, the watermark model fidelity results in a marginal decrease of 0.05 in the IS for CIFAR-10 and 0.01 for CelebA compared with the fine-tuned baseline model.
Additionally, there is an average slight increase in the FID by 0.6 for CIFAR-10 and 0.1 for CelebA on average.
Compared to mWDM setting, sWDM setting results in less degradation of the model fidelity.
This difference can be caused by the varying amount of information contained within the embedded watermark data, which is called the watermark capacity.
The results indicate that the WDM setting with a larger watermark capacity tends to exhibit greater degradation in model fidelity.

\begin{figure}[htbp]
   \centering
   \includegraphics[width=0.47\textwidth]{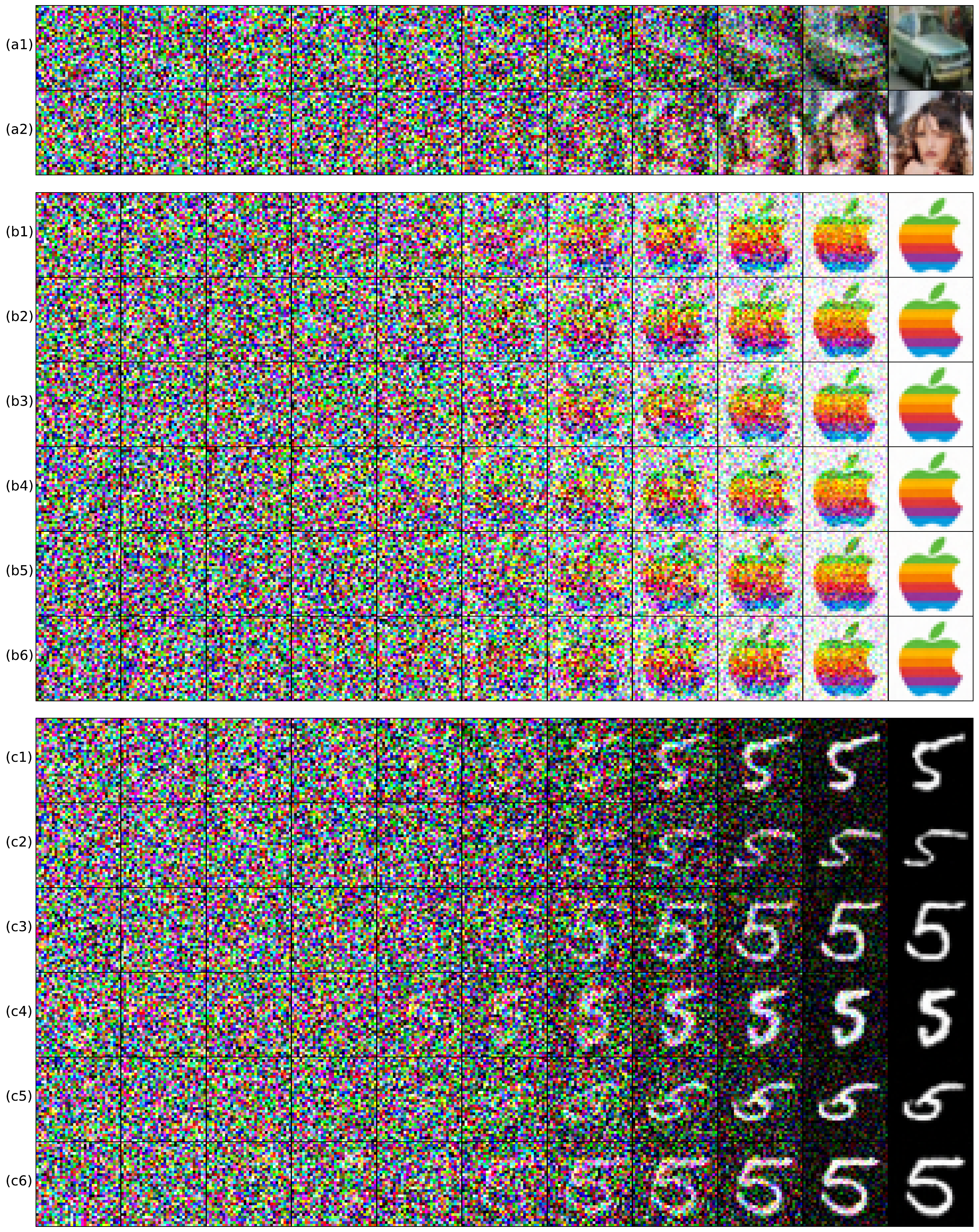}
    \caption{Illustration of task and watermark sampling processes. 
    The subfigures (a1) and (a2) illustrate the task sampling process of the watermarked model in CIFAR-10 and CelebA, respectively, under the sWDM-sel setting.
    The subfigures (b1) to (b3) illustrate the watermark sampling process in CIFAR-10 under sWDM-sel, sWDM-randp, and sWDM-randc settings, respectively, and similar for (b4) to (b6) in CelebA.
    The subfigures (c1) to (c3) illustrate the watermark sampling process in CIFAR-10 under mWDM-sel, mWDM-randp, and mWDM-randc settings, respectively, and similar for (c4) to (c6) in CelebA.}
    \label{fig: process_demo}
    \vspace{-1em}
\end{figure}

We demonstrate the task and watermark sampling processes in \cref{fig: process_demo}.
It can be observed that the embedded watermark does not affect the task generation results or get exposed during the task sampling.
The watermark data can only be retrieved using the specific trigger provided during the embedding process.
The watermark fidelity in \cref{tab: model_fidelity} is evaluated based on the watermark similarity (denoted as WS) and verification confidence level in hypothesis testing (denoted as VC).
We observe that the watermark similarities evaluated using SSIM in all sWDM settings are significantly higher than those of the reference baseline models, which contain no embedded watermark.
A similar trend can be found in all mWDM settings, where the similarities measured in FID are notably lower than those of the baseline models.
The verification confidence levels, indicated by the $p$-values in hypothesis testing of all WDM settings, are also notably lower than the significance level $\alpha=0.01$.
It indicates that the null hypothesis $H_0:\boldsymbol{\mu}_{c} \leq \boldsymbol{\mu}_{s}$, which shows the suspected model is not stolen, can be rejected with a high level of statistical significance.
These outcomes demonstrate that all WDM settings exhibit high watermark fidelity, which implies that the embedded watermark can be effectively extracted from the watermarked model and verified.

\subsection{Watermark Robustness}
We evaluate the robustness of our proposed WDM against three main kinds of removing attacks, which are 1) model compression, 2) weight perturbation, and 3) model fine-tuning.
\subsubsection{Model Compression}
We select the widely-used model quantization technique as our evaluated model compression attack.
In quantization, the model parameters are typically stored using numbers of lower numerical precision. Practically, we convert the model parameters from 32-bit floating-point numbers to 16-bit floating-point numbers, which reduces the model's size by nearly half.
\cref{tab: quantization} displays the watermark fidelity of watermarked models reduced by quantization under various WDM settings.
We observe that the watermark similarity exhibits no significant loss after quantization.
The $|\delta\text{WS}|$ for all sWDM settings is less than $10^{-4}$, while for all mWDM settings, it is about 0.2.
These changes are relatively small compared to the results presented in \cref{tab: model_fidelity}.
Similarly, the verification confidence level also remains nearly unchanged, with all $p$-values falling below the significance level of $10^{-3}$.
These results demonstrate that the WDM is robust against model quantization attacks in all six evaluated settings.
\begin{table}[htbp]
  \centering
\setlength{\tabcolsep}{2pt}
\scalebox{0.88}{
\begin{tabular}{@{}l|ccc|ccc@{}}
\toprule
 & \multicolumn{3}{c|}{CIFAR-10} & \multicolumn{3}{c}{CelebA} \\ \midrule
Methods & WS & $|\delta\text{WS}|$ & VC$\downarrow$ & WS & $|\delta\text{WS}|$ & VC$\downarrow$ \\ \midrule
%sWDM-ref & 0.8082988903806431 & $10^{-6}$ & $<10^{-324}$ & 0.8059294741667316 & $10^{-6}$ & $<10^{-324}$ \\
sWDM-sel & 0.997 & $10^{-6}$ & $<\!10^{-324}$ & 0.998 & $10^{-6}$ & $<\!10^{-324}$ \\
sWDM-randp & 0.997 & $10^{-5}$ & $<\!10^{-324}$ & \bf{0.998} & $10^{-4}$ & $\bf{<\!10^{-324}}$ \\
sWDM-randc & \bf{0.998} & $\bf{10^{-6}}$ & $\bf{<\!10^{-324}}$ & 0.998 & $10^{-6}$ & $<\!10^{-324}$ \\
sWDM-avg & 0.997 & $10^{-6}$ & $<\!10^{-324}$ & 0.998 & $\bf{10^{-5}}$ & $<\!10^{-324}$ \\ \midrule
%mWDM-ref & 41.94 & - & - & 43.89 & - & - \\
mWDM-sel & 6.81 & 0.034 & $10^{-28}$ & 8.76 & \bf{0.045} & $10^{-28}$ \\
mWDM-randp & 7.39 & \bf{0.025} & $10^{-28}$ & 8.14 & 0.226 & $10^{-28}$ \\
mWDM-randc & \bf{6.00} & 0.093 & $\bf{10^{-28}}$ & \bf{5.79} & 0.155 & $\bf{10^{-28}}$ \\
mWDM-avg & 6.73 & 0.034 & $10^{-28}$ & 7.56 & 0.112 & $10^{-28}$ \\ \bottomrule
\end{tabular}
}
  \caption{Watermark fidelity under different WDM settings after the quantization attack.
  The same notion in \cref{tab: model_fidelity} is used for the numeric results that are equal to 0, denoted by $<\!10^{-324}$.
  The best values are highlighted in bold.}
  \label{tab: quantization}
  \vspace{-1em}
\end{table}

\subsubsection{Weight Perturbation}
\begin{figure}[htbp]
  \centering
   \includegraphics[width=0.47\textwidth]{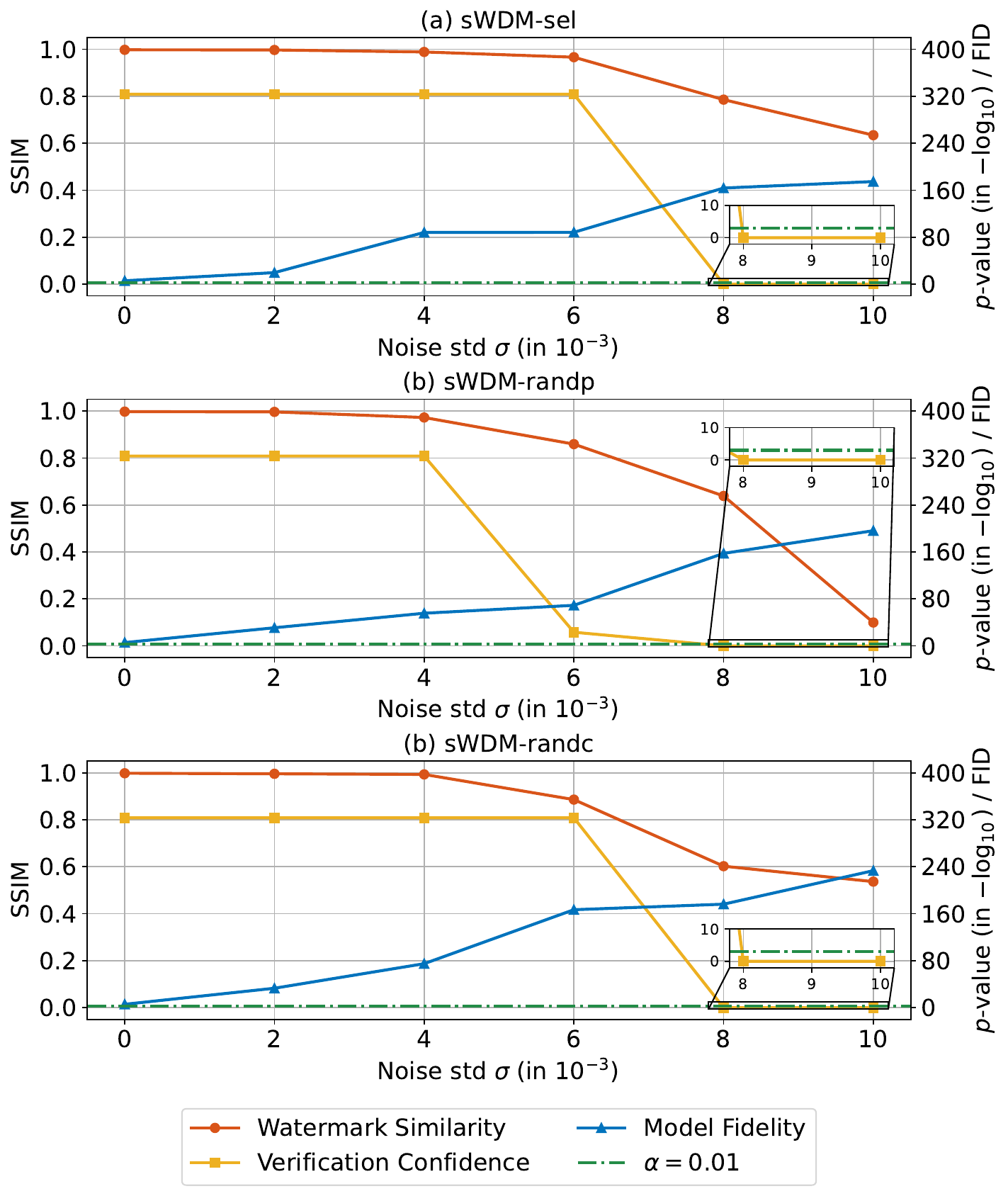}
   \caption{Model and watermark fidelity of watermarked models after the weight perturbation attack under sWDM-sel, sWDM-randp, and sWDM-randc settings in CIFAR-10. The $p$-value is shown in $-\log_{10}$.}
   \label{fig: weight_perturbation}
   \vspace{-1em}
\end{figure}
Model weight perturbation is a commonly used method for removing model watermarks. 
This attack adds random noises to the model weights, which perturbs the activation of the hidden embedded watermark while preserving the model fidelity.
In \cref{fig: weight_perturbation}, the changes in both model and watermark fidelity are illustrated across different standard deviations for the added noise under the sWDM-sel, sWDM-randp, and sWDM-randc settings in CIFAR-10.
We observe that for all three WDM settings, the embedded watermark can still be extracted and verified with or before $\sigma = 6\!\times\!10^{-3}$.
Subsequently, when the standard deviation of added noise goes above this threshold, the model fidelity in FID exceeds 100, indicating that the model's functionality for the original task is severely damaged.
The verification confidence level before this threshold is lower than the significance level $\alpha=0.01$.
These results demonstrate that the evaluated WDM settings are robust against the model weight perturbation attack, with the task performance remaining acceptable.

\subsubsection{Model Fine-tuning}
\begin{figure}[htbp]
  \centering
   \includegraphics[width=0.47\textwidth]{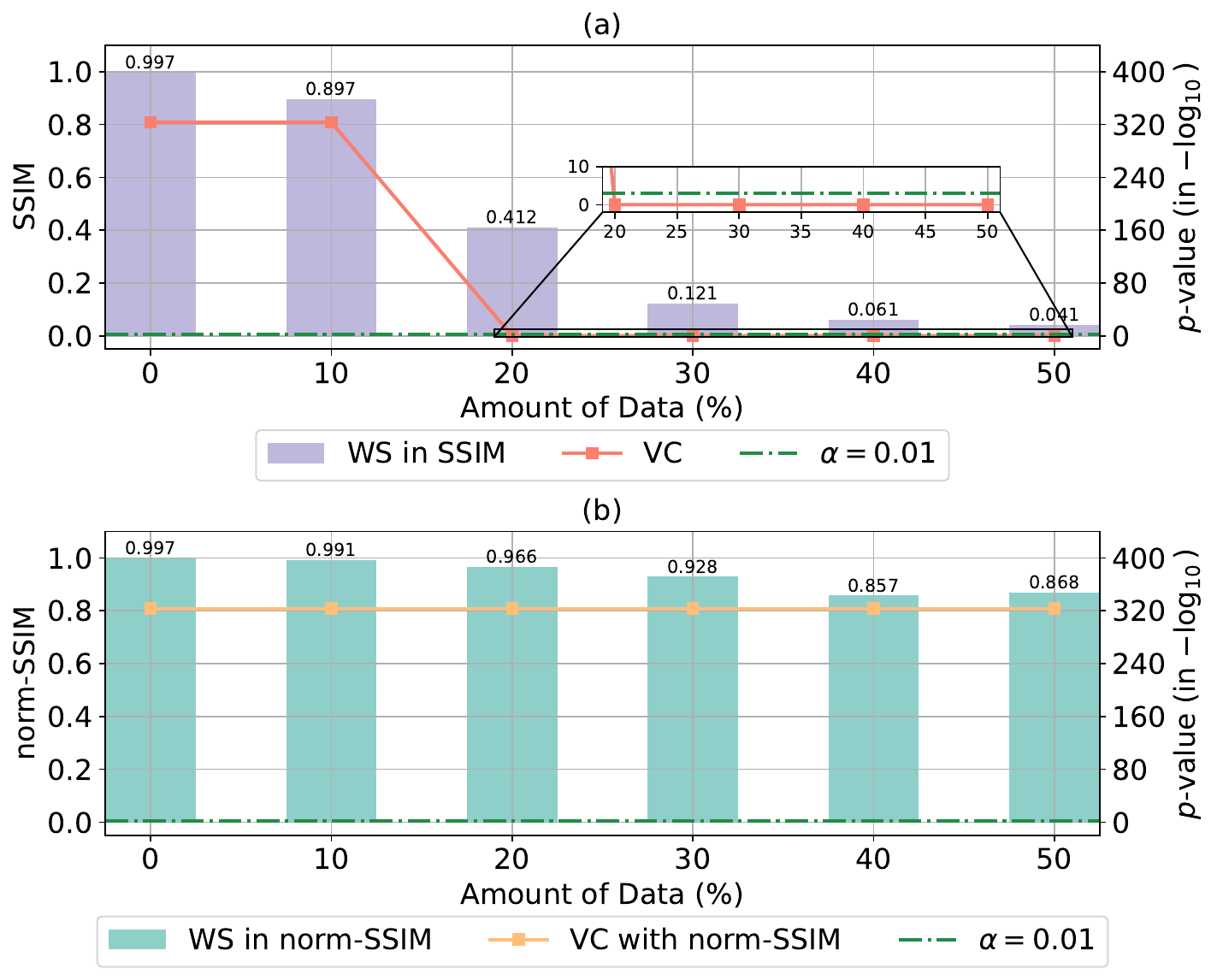}
   \caption{The watermark fidelity and verification confidence level of the sWDM-sel setting in CIFAR-10 after the model fine-tuning attacks. Subfigure (a) represents the WS and VC computed using SSIM and subfigure(b) represents those computed using norm-SSIM.}
   \label{fig: fine-tuning}
   \vspace{-1em}
\end{figure}

Model fine-tuning is another commonly used attack for removing embedded watermarks.
We evaluate fine-tuning all layers of the watermarked model using the same data sampled from its training dataset.
The change of watermark fidelity, verification confidence level and the data amount used for watermark embedding under sWDM-sel setting in CIFAR-10 are illustrated in \cref{fig: fine-tuning}.
In subfigure (a), we observe the watermark fidelity of the sWDM-sel setting falls after using 20\% amount of data during watermark embedding.
The verification confidence level also gets higher than the significance level $\alpha$, indicating that the watermark is removed after using 20\% % of data during embedding.
However, if we take the norm-SSIM, which only calculates the mean of SSIM values larger than 0.8, as the watermark similarity, the watermark can remain when using 50\% amount of data during embedding.
The results indicate that our proposed WDM method is only robust against the removing attack of fine-tuning all layers using a small amount of data during the watermark embedding.
The possible reason is that fine-tuning all layers of the diffusion models with a large amount of data may cause catastrophic forgetting since the fine-tuning optimization objective is not included with the watermark embedding part.

\subsection{Ablation \& Supplementary Experiments}
% \subsubsection{Impact of the Timestep in Sampling}
% \begin{figure}[t]
%   \centering
%    \includegraphics[width=0.5\textwidth]{figures/sWDM-sel-wp.pdf}
%    \caption{Overwriting.}
%    \label{fig:onecol}
% \end{figure}
% \subsubsection{Impact of Trigger Factor $\gamma_1$}
Due to the page limit, we put the ablation study and supplementary experiments in Sec.10 in Appendix.

%================================================= 
\section{Conclusions}
%================================================= 
In this paper, we propose WDM, a complete watermarking framework for diffusion models that consists of watermark embedding, extraction, and verification.
The watermark data is embedded through concurrently learning a WDP and a standard diffusion process for task data generation.
Watermark extraction can only be processed by obtaining the original trigger, which does not expose the watermark during task generation and saves the watermark imperceptibility.
The extracted watermark can be verified by similarity comparison and hypothesis testing, giving the conclusion regarding the model ownership.
We present the detailed theoretical foundations and analysis of the used WDP and validate the effectiveness of our proposed WDM through experiments.
We evaluate the robustness of WDM in various settings against model compression, weight perturbation attacks, and model fine-tuning attacks.
Overall, the proposed WDM can achieve the goals of model fidelity, watermark fidelity, and watermark robustness. 
Introducing this new approach has provided a powerful tool for protecting intellectual property in diffusion models.
{
    \small
    \bibliographystyle{ieeenat_fullname}
    \bibliography{main}
}

% WARNING: do not forget to delete the supplementary pages from your submission 
\clearpage
\setcounter{page}{1}
\maketitlesupplementary

%\begin{strip}
\section{Theoretical Foundations for the MDP and the WDP}
\label{appendix: proof}
%=================================================
\subsection{Proof of \cref{theorem: mdp}}
This subsection aims to prove\cref{theorem: mdp}.
Consider a scenario where $t\in\{1,...,T\}$.
In this context, the forward transition kernel of a MDP can be represented as follows:
\begin{equation}
    q(\mathbf{x}_t|\mathbf{x}_{t\!-\!1})\sim\mathcal{N}(\mathbf{x}_t;\sqrt{\alpha_t}\mathbf{x}_{t\!-\!1}\!+\!\sqrt{\alpha_t}\cone_t,\ctwo^{2}(1\!-\!\alpha_t)\mathbf{I})
\end{equation}
   
\begin{proof}
By applying the re-parameterization trick, we can get the following result:
\begin{align}
    &\mathbf{x}_t=\sqrt{\alpha_t}\mathbf{x}_{t\!-\!1}\!+\!\sqrt{\alpha_t}\cone_t\!+\!\ctwo\sqrt{1\!-\!\alpha_t}\boldsymbol{\epsilon}_{t}\\
    &=\sqrt{\alpha_t}(\sqrt{\alpha_{t\!-\!1}}\mathbf{x}_{t\!-\!2}\!+\!\sqrt{\alpha_{t\!-\!1}}\cone_{t\!-\!1}\notag\\
    &\quad\!+\!\ctwo\sqrt{1\!-\!\alpha_{t\!-\!1}}\boldsymbol{\epsilon}_{t\!-\!1})\!+\!\sqrt{\alpha_t}\cone_{t}\!+\!\ctwo\sqrt{1\!-\!\alpha_t}\boldsymbol{\epsilon}_{t}\\
    &=\sqrt{\alpha_t\alpha_{t\!-\!1}}\mathbf{x}_{t\!-\!2}\!+\!(\sqrt{\alpha_t}\cone_t\!+\!\sqrt{\alpha_t\alpha_{t\!-\!1}}\cone_{t\!-\!1})
    \notag\\
    &\quad\!+\!\ctwo(\sqrt{\alpha_{t}(1\!-\!\alpha_{t\!-\!1})}\boldsymbol{\epsilon}_{t\!-\!1}\!+\!\sqrt{1\!-\!\alpha_t}\boldsymbol{\epsilon}_{t})\\
    &=\sqrt{\alpha_t\alpha_{t\!-\!1}}\mathbf{x}_{t\!-\!2}\!+\!(\sqrt{\alpha_t}\cone_t\!+\!\sqrt{\alpha_t\alpha_{t\!-\!1}}\cone_{t\!-\!1})\notag\\
    &\quad\!+\!\ctwo\sqrt{1\!-\!\alpha_{t}\alpha_{t\!-\!1}}\boldsymbol{\epsilon}_{t,t\!-\!1}\\
    &=\sqrt{\Bar{\alpha}_t}\mathbf{x}_{0}\!+\!\sum_{i=1}^{t}(\sqrt{\prod_{j={t\!+\!1\!-\!i}}^{t}\alpha_j}\cone_{t\!+\!1\!-\!i})\!+\!\ctwo\sqrt{1\!-\!\Bar{\alpha}_{t}}\hat{\boldsymbol{\epsilon}},
\end{align}
where $\boldsymbol{\epsilon}_{t,t\!-\!1},\hat{\boldsymbol{\epsilon}} \sim \mathcal{N}(0,1)$.
We can observe that the distribution of $ q(\mathbf{x}_{t}|\mathbf{x}_{0})$ is still a Gaussian distribution and can be implicitly represented as follows:
\begin{equation}
    q(\mathbf{x}_{t}|\mathbf{x}_{0})\sim \mathcal{N}(\mathbf{x}_{0};\sqrt{\Bar{\alpha}_t}\mathbf{x}_{0}\!+\!\Tilde{\cone}_t,\ctwo^{2}(1\!-\!\Bar{\alpha}_t)\mathbf{I}),
\end{equation}
where 
\begin{equation}
\label{eq: tiddle_phi_t}
    \Tilde{\cone}_t=\sum_{i=1}^{t}(\sqrt{\prod_{j={t\!+\!1\!-\!i}}^{t}\alpha_j}\cone_{t\!+\!1\!-\!i}).
\end{equation}
The transition kernel conditioned on $\mathbf{x}_0$ in the reverse process of an MDP can be represented as follows:
\begin{equation}
    q(\mathbf{x}_{t\!-\!1} \vert \mathbf{x}_t, \mathbf{x}_0) = \mathcal{N}(\mathbf{x}_{t\!-\!1}; \tilde{\boldsymbol{\mu}}_t(\mathbf{x}_t, \mathbf{x}_0), \tilde{\sigma}_t \mathbf{I}).
\end{equation}
By applying the Bayes rule, we obtain the following result:
{\small
\begin{align}
&q(\mathbf{x}_{t\!-\!1} \vert \mathbf{x}_t, \mathbf{x}_0) = q(\mathbf{x}_t \vert \mathbf{x}_{t\!-\!1}, \mathbf{x}_0) \frac{ q(\mathbf{x}_{t\!-\!1} \vert \mathbf{x}_0) }{ q(\mathbf{x}_t \vert \mathbf{x}_0) } \\
&\propto \!\exp \!\Big(\!-\!\frac{1}{2} \big(\frac{(\mathbf{x}_t\!-\!\sqrt{\alpha_t} \mathbf{x}_{t\!-\!1}-\sqrt{\alpha_t}\cone_t)^2}{\ctwo^2(1\!-\!\alpha_t)} \notag\\
&\quad\!+\! \frac{(\mathbf{x}_{t\!-\!1}\!-\!\sqrt{\bar{\alpha}_{t\!-\!1}} \mathbf{x}_0-\Tilde{\cone}_{t\!-\!1})^2}{\ctwo^2(1\!-\!\bar{\alpha}_{t\!-\!1})} \notag\\
&\quad- \frac{(\mathbf{x}_t\!-\!\sqrt{\bar{\alpha}_t} \mathbf{x}_0-\Tilde{\cone}_{t})^2}{\ctwo^2(1\!-\!\bar{\alpha}_t)} \big) \Big) \\
&= \exp \! \Big(\! -\!\frac{1}{2\ctwo^2} \big(\frac{(\mathbf{x}_t\!-\!\sqrt{\alpha_t}\cone_t)^2 \!-\! 2\sqrt{\alpha_t} (\mathbf{x}_t\!-\!\sqrt{\alpha_t}\cone_t) \mathbf{x}_{t\!-\!1} \!+\! \alpha_t \mathbf{x}_{t\!-\!1}^2} {1\!-\!\alpha_t} \notag\\
&\quad\!+\! \frac{ \mathbf{x}_{t\!-\!1}^2\!-\!2 (\sqrt{\bar{\alpha}_{t\!-\!1}}\mathbf{x}_{0}\!+\!\Tilde{\cone}_{t\!-\!1})\mathbf{x}_{t\!-\!1}\!+\!( \sqrt{\bar{\alpha}_{t\!-\!1}}\mathbf{x}_0 \!+\!\Tilde{\cone}_t)^2}  {1\!-\!\bar{\alpha}_{t\!-\!1}}\notag\\
&\quad- \frac{(\mathbf{x}_t\!-\!\sqrt{\bar{\alpha}_{t\!-\!1}} \mathbf{x}_0-\Tilde{\cone}_t)^2}{1\!-\!\bar{\alpha}_t} \big) \Big) \\
&= \exp\Big( -\frac{1}{2} \big( \frac{1}{\ctwo^2}(\frac{\alpha_t}{1\!-\!\alpha_t} \!+\! \frac{1}{1\!-\!\bar{\alpha}_{t\!-\!1}}) \mathbf{x}_{t\!-\!1}^2\!-\!\frac{2}{\ctwo^2}(\frac{\sqrt{\alpha_t}(\mathbf{x}_t\!-\!\sqrt{\alpha_t}\cone_t)}{1\!-\!\alpha_t} \notag\\
&\quad\!+\! \frac{\sqrt{\bar{\alpha}_{t\!-\!1}}\mathbf{x}_0\!+\!\Tilde{\cone}_{t\!-\!1}}{1\!-\!\bar{\alpha}_{t\!-\!1}} )\mathbf{x}_{t\!-\!1} \color{black}{C(\mathbf{x}_t, \mathbf{x}_0) \big) \Big)}.
\end{align}
}
We can also derive the mean and variance of this Gaussian distribution as follows:
\begin{equation}
    \tilde{\sigma}_t=\frac{1}{\frac{1}{\ctwo^2}(\frac{\alpha_t}{1\!-\!\alpha_t} \!+\! \frac{1}{1\!-\!\bar{\alpha}_{t\!-\!1}})}=\frac{\ctwo^2(1\!-\!\bar{\alpha}_{t\!-\!1})(1\!-\!\alpha_t)}{1\!-\!\bar{\alpha}_t},
\end{equation}
\begin{align}
&\tilde{\boldsymbol{\mu}}_t (\mathbf{x}_t, \mathbf{x}_0)= \frac{1}{\ctwo^2}(\frac{\sqrt{\alpha_t}(\mathbf{x}_t\!-\!\sqrt{\alpha_t}\cone_t)}{1\!-\!\alpha_t}  \!+\! \frac{\sqrt{\bar{\alpha}_{t\!-\!1}}\mathbf{x}_0\!+\!\Tilde{\cone}_{t\!-\!1}}{1\!-\!\bar{\alpha}_{t\!-\!1}} )\notag\\
&\quad/\frac{1}{\ctwo^2}(\frac{\alpha_t}{1\!-\!\alpha_t} \!+\! \frac{1}{1\!-\!\bar{\alpha}_{t\!-\!1}})  \\
&=\frac{\sqrt{\alpha_t}(1\!-\!\bar{\alpha}_{t\!-\!1})}{1\!-\!\bar{\alpha}_t} \mathbf{x}_t \!+\! \frac{\sqrt{\bar{\alpha}_{t\!-\!1}}(1\!-\!\alpha_t)}{1\!-\!\bar{\alpha}_t} \mathbf{x}_0\notag\\
&\quad+\!\frac{\bar{\alpha}_t\!-\!\alpha_t}{1\!-\!\bar{\alpha}_t}\cone_t\!+\!\frac{1\!-\!\alpha_t}{1\!-\!\bar{\alpha}_t}\Tilde{\cone}_{t\!-\!1}.
\end{align}
% t>1
By substituting $\mathbf{x}_0$ above with 
\begin{equation}
    \mathbf{x}_0 = \frac{1}{\sqrt{\bar{\alpha}_t}}(\mathbf{x}_t\!-\!\Tilde{\cone}_{t}-\ctwo\sqrt{1\!-\!\bar{\alpha}_t}\boldsymbol{\epsilon}_t),
\end{equation}
we obtain the following result:
\begin{equation}
\label{mdp_mean}
    \tilde{\boldsymbol{\mu}}_t=\frac{1}{\sqrt{\alpha_t}} \Big( \mathbf{x}_t\!-\!\ctwo\frac{1\!-\!\alpha_t}{\sqrt{1\!-\!\bar{\alpha}_t}} \boldsymbol{\epsilon}_t\Big)\!+\!C_t,
\end{equation}
where $C_t$ is a constant term.
Following the same simplification procedure as demonstrated in the DDPM~\cite{ho2020denoising}, we can represent the optimization objective during training as follows:
\begin{align}
&L^{\text{\mdp}}_{t}= \mathbb{E}_{t \sim [1, T], \mathbf{x}_0, \boldsymbol{\epsilon}_t} \Big[\|\boldsymbol{\epsilon}_t\!-\!\boldsymbol{\epsilon}_\theta(\mathbf{x}_t, t)\|^2 \Big] \\
&= \mathbb{E}_{t \sim [1, T], \mathbf{x}_0, \boldsymbol{\epsilon}_t} \Big[\|\boldsymbol{\epsilon}_t\!-\!\boldsymbol{\epsilon}_\theta(\sqrt{\bar{\alpha}_t}\mathbf{x}_0 \!+\! \sqrt{1\!-\!\bar{\alpha}_t}\boldsymbol{\epsilon}_t, t)\|^2 \Big]
\end{align}
The simplified training loss shares a similar form as the one in the DDPM. 
Then we complete the proof.
\end{proof}

\subsection{Proof of \cref{theorem: wdp}}
\begin{proof}
By replacing the $\mathbf{x}_{t}$ in \cref{eq: wdp} using \cref{eq: sdp_xt}, we can get 
\begin{align}
     \Tilde{\mathbf{x}}_{t}&=\gamma_1{\mathbf{x}_{t}}\!+\!(1\!-\!\gamma_1)\mathbf{b}\\
     &=\gamma_1(\sqrt{\Bar{\alpha}_t}\mathbf{x}_0\!+\!\sqrt{1\!-\!\Bar{\alpha}_t}\boldsymbol{\epsilon})\!+\!(1\!-\!\gamma_1)\mathbf{b}.
\end{align}
Then we replace the $\mathbf{x}_0$ using \cref{eq: sdp_xt} to get
\begin{align}
     \Tilde{\mathbf{x}}_{t}&=\gamma_1(\sqrt{\Bar{\alpha}_t}\mathbf{x}_0\!+\!\sqrt{1\!-\!\Bar{\alpha}_t}\boldsymbol{\epsilon})\!+\!(1\!-\!\gamma_1)\mathbf{b}\\
     &=\gamma_1\sqrt{\Bar{\alpha}_t}(\frac{\Tilde{\mathbf{x}}_{0}-(1\!-\!\gamma_1)\mathbf{b}}{\gamma_1})\!+\!\gamma_1\sqrt{1\!-\!\Bar{\alpha}_t}\boldsymbol{\epsilon}\!+\!(1\!-\!\gamma_1)\mathbf{b}\\
     &=\sqrt{\Bar{\alpha}_t}\Tilde{\mathbf{x}}_0\!+\!(1\!-\!\sqrt{\Bar{\alpha}_t})(1\!-\!\gamma_1)\mathbf{b}\!+\!\gamma_1\sqrt{1\!-\!\Bar{\alpha}_t}\boldsymbol{\epsilon}.
\end{align}
Compare $q(\Tilde{\mathbf{x}}_t|\Tilde{\mathbf{x}}_{0})$ to the diffusion process with the modified Gaussian kernel, we can get they are the same if the configurations satisfy
\begin{equation}
    \Tilde{\cone}_t=(1\!-\!\sqrt{\Bar{\alpha}_t})(1\!-\!\gamma_1)\mathbf{b}
\end{equation}
and
\begin{equation}
    \ctwo=\gamma_1.
\end{equation}
Based on \cref{eq: tiddle_phi_t}, we can represent $\cone_t$ as 
\begin{equation}
    \cone_t=\frac{\Tilde{\cone}_t}{\sqrt{\alpha_t}}-\Tilde{\cone}_{t\!-\!1}=(\frac{1}{\sqrt{\alpha_t}}-1)(1\!-\!\gamma_1)\mathbf{b}
\end{equation}
Then we complete the proof.
\end{proof}
%\end{strip}

%=================================================
\section{Trigger and Trigger Factor Selection}
\label{appendix: trigger_selection}
%=================================================
To guarantee that the watermark is not exposed during task generation, it is essential to carefully choose the trigger $\mathbf{b}$ and trigger factor $\gamma_1$.
The selected $\mathbf{b}$ and $\gamma_1$ should create enough divergence between the state distribution of the watermark diffusion process and the standard diffusion process, which can prevent the leakage of the watermark.
We denote the state distribution in watermark extraction as $q(\mathbf{x}^{w}_{t\!-\!1} \vert \mathbf{x}^{w}_t, \mathbf{x}^{w}_0)$.
We also denote the state distribution in the standard reverse diffusion process as $q(\tilde{\mathbf{x}}^{w}_{t\!-\!1} \vert \tilde{\mathbf{x}}^{w}_t, \tilde{\mathbf{x}}^{w}_0)$. 
It is necessary to ensure that the divergence between these two state distributions is sufficient, which guarantees that the watermark data are not extracted during the task generation.
In the case where there is no divergence between the two distributions, such as when 
$\gamma_1=1$ or $\mathbf{b}=\mathbf{0}$, the WDP transforms into the standard diffusion process for the watermark data. 
This overlap can consequently lead to the leakage of the watermark.
By substituting the $\tilde{\cone}_t$ and $\ctwo$ in \cref{mdp_mean} with \cref{eq: cwt}, we can derive the mean value of the WDP state distribution as follows:
\begin{equation}
\tilde{\boldsymbol{\mu}}_t (\tilde{\mathbf{x}}^{w}_t)\!=\!\frac{1}{\sqrt{\alpha_t}} \Big( \tilde{\mathbf{x}}^{w}_t\!-\!\gamma_1\frac{1\!-\!\alpha_t}{\sqrt{1\!-\!\bar{\alpha}_t}} \boldsymbol{\epsilon}_t\Big)\!\!+\!\!(1\!-\!\frac{1}{\sqrt{\alpha_t}})(1\!-\!\gamma_1)\mathbf{b}.
\end{equation}
Since the mean value of $\mathbf{x}^{w}_{t\!-\!1}$ in the standard reverse process is as follows:
\begin{equation}
     \boldsymbol{\mu}_t(\mathbf{x}^{w}_t)=\frac{1}{\sqrt{\alpha_t}}(\mathbf{x}^{w}_t\!-\!\frac{1\!-\!\alpha_t}{\sqrt{1\!-\!\Bar{\alpha}_t}}\boldsymbol{\epsilon}_t).
\end{equation}
We can derive their difference as follows:
\begin{equation}
     \tilde{\boldsymbol{\mu}}_t\!-\!\boldsymbol{\mu}_t=\frac{1}{\sqrt{\alpha_t}}(\mathbf{x}_t\!-\!\frac{1\!-\!\alpha_t}{\sqrt{1\!-\!\bar{{\alpha}_t}}}\boldsymbol{\epsilon}_t\!-\!\mathbf{b})(1\!-\!\gamma_1).
\end{equation}
The trigger $\mathbf{b}$ needs to be chosen from a data distribution distinct from the one encountered in the standard diffusion process for task data.
In practice, we choose an Out\!-\!Of-Distribution (OOD) sample from the task data, such as a selected image with ownership information as the trigger, as we have done in our experimental settings.

%=================================================
\section{Experiment Implementation Details}
\label{appendix: exp_implementation}
%=================================================
\subsection{Network Implementation and Parameters}
Our implementation of DDPMs is based on the approach presented in~\cite{nichol2021improved}, which utilizes the U-Net~\cite{unet} architecture introduced in~\cite{ho2020denoising} as the backbone model. 
The U-Net model architecture utilizes stacked residual layers, complemented by both downsampling and upsampling convolutions. Each timestep is encoded through positional encoding, which is then integrated into every residual block. 
This integration is achieved using a global attention layer with multiple heads for enhanced processing capability.
In our experimental setup, we configured the global attention layer with four heads and set its resolution to $16\times 16$ and $8\times 8$.
Additionally, we set the base number of channels at 128 and incorporated three residual blocks at each resolution level.

\subsection{Training and Fine-tuning Parameters}
To obtain the baseline diffusion models on CIFAR-10 and CelebA datasets, we followed the hyper-parameter settings outlined in~\cite{nichol2021improved} and set the image size to $32\times 32$ pixels. 
We trained each model independently at a learning rate of $10^{-4}$ over 300K batch iterations, with each iteration handling a data batch of size 128. 
Additionally, we employed the Exponential Moving Average (EMA) model for sampling, setting the EMA rate at 0.999.
We fine-tune the baseline model for an additional 20K iterations for all WDM settings to embed the watermark with the same batch size as 128.
The trigger factor is set to  $\gamma_1=0.8$, and the trade-off factor $\gamma_2$ during fine-tuning is set to 0.1 for sWDM settings and 0.2 for mWDM settings.

%=================================================
\section{Supplementary Experiments}
\label{appendix: supply_exps}
%=================================================
\subsection{Impact of Watermark Extraction Timestep}
\begin{figure}[htbp]
\begin{center}
\includegraphics[width=0.47\textwidth]{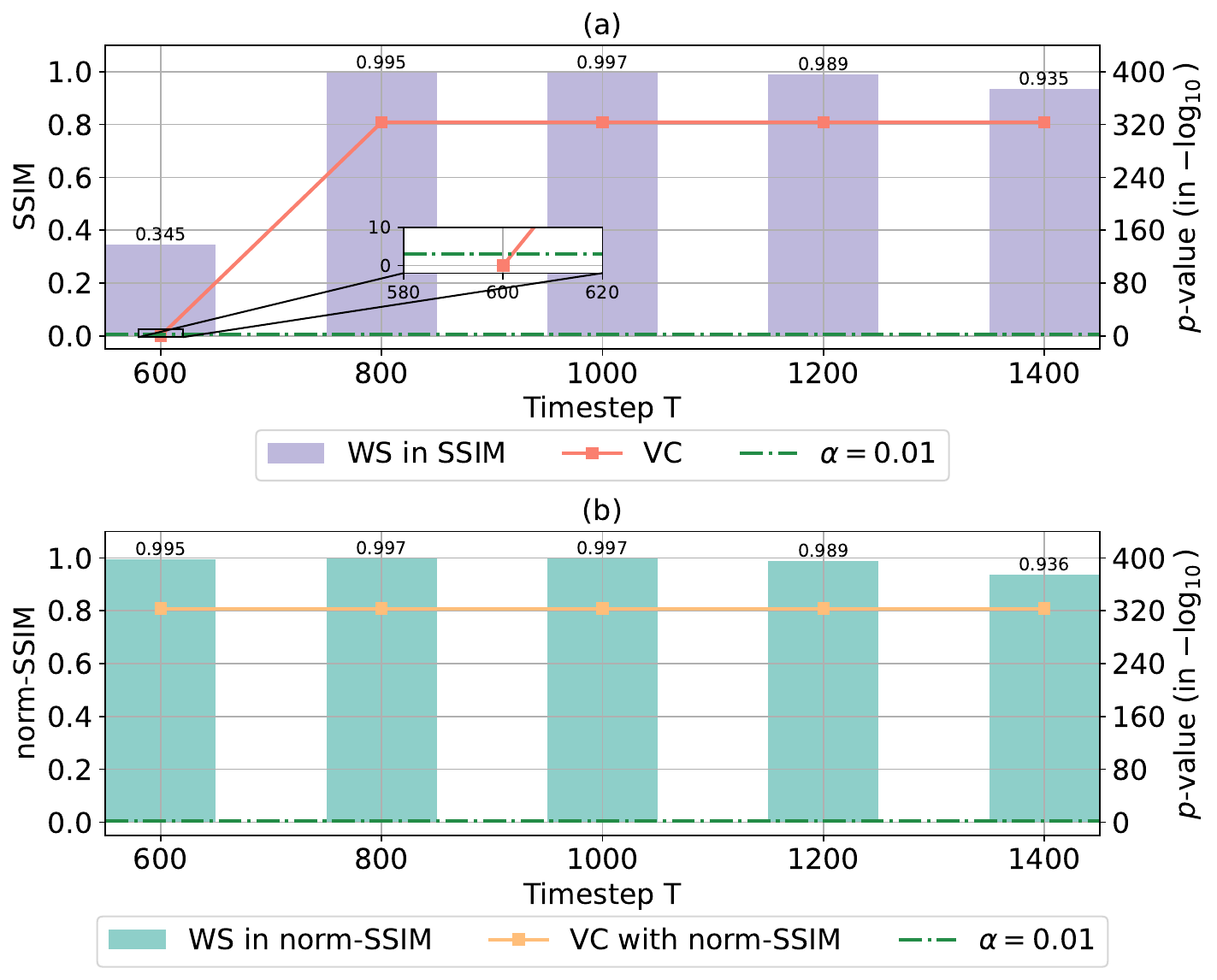}
\end{center}
\caption{The watermark fidelity and verification confidence level of the sWDM-sel setting on CIFAR-10 with different timesteps during watermark extraction. Subfigure (a) represents the WS and VC computed using SSIM and subfigure(b) represents those computed using norm-SSIM.}
\label{fig: ab_ts}
\end{figure}
While the timestep used in the watermark extraction process can also be considered a part of the secret, we still evaluate the impact of that on the watermark fidelity.
The results from this evaluation are illustrated in \autoref{fig: ab_ts}.
When evaluating watermark verification confidence based on SSIM watermark similarity, we can observe that the confidence level remains above the significance threshold for timesteps $t\geq 800$.
However, upon employing norm-SSIM as the evaluation metric, the verification confidence consistently exceeds the significance level across all tested timesteps, including at $t=600$.
This observation indicates that the timestep used in watermark extraction does not significantly affect the effectiveness of our proposed WDM.

\subsection{Watermark Extraction in DDIM~\cite{song2020denoising}}
\begin{figure}[htbp]
\begin{center}
\includegraphics[width=0.47\textwidth]{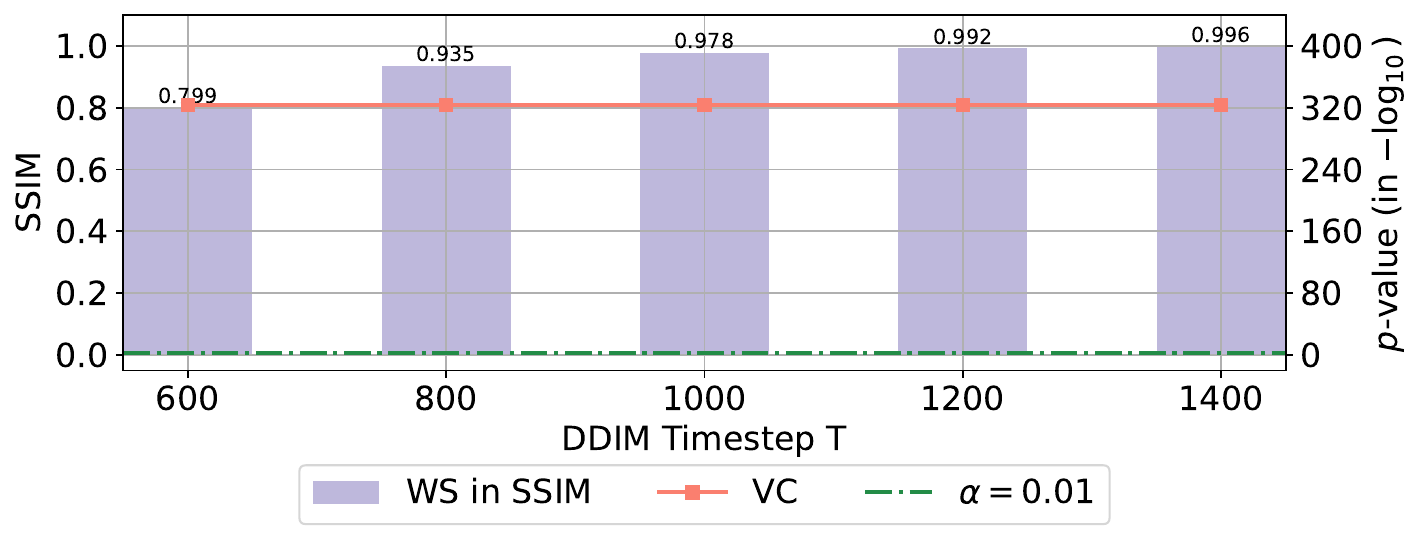}
\end{center}
\caption{The watermark fidelity and verification confidence level of the sWDM-sel setting on CIFAR-10 with different DDIM timesteps during watermark extraction.}
\label{fig: ab_ddim}
\end{figure}

We further evaluate the WDM settings with various timesteps in the watermark extraction process utilizing the DDIM architecture~\cite{song2020denoising} instead of the DDPM.
DDIM architecture modifies the diffusion process, allowing for a deterministic and potentially reduced number of steps in the reverse diffusion. 
The results are illustrated in \autoref{fig: ab_ddim}.
We can observe that verification confidence levels for all evaluated timesteps in DDIM remain above the significance threshold, 
This indicates that employing the DDIM architecture instead of the DDPM does not significantly impact the effectiveness of our proposed WDM.

\subsection{Visual Samples of Baseline and Watermark Models}
\begin{figure}[htbp]
\begin{center}
\includegraphics[width=0.47\textwidth]{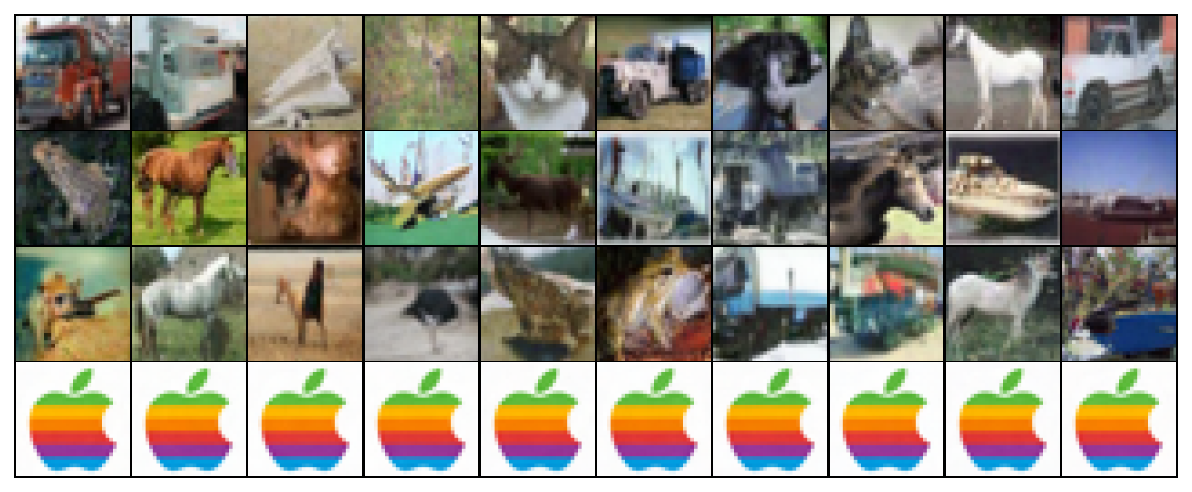}
\end{center}
\caption{Visual examples sampled from the baseline model for sWDM-sel on CIFAR-10.
The above three lines are task generation results and the bottom line are watermark extraction results.}
\end{figure}

\begin{figure}[htbp]
\begin{center}
\includegraphics[width=0.47\textwidth]{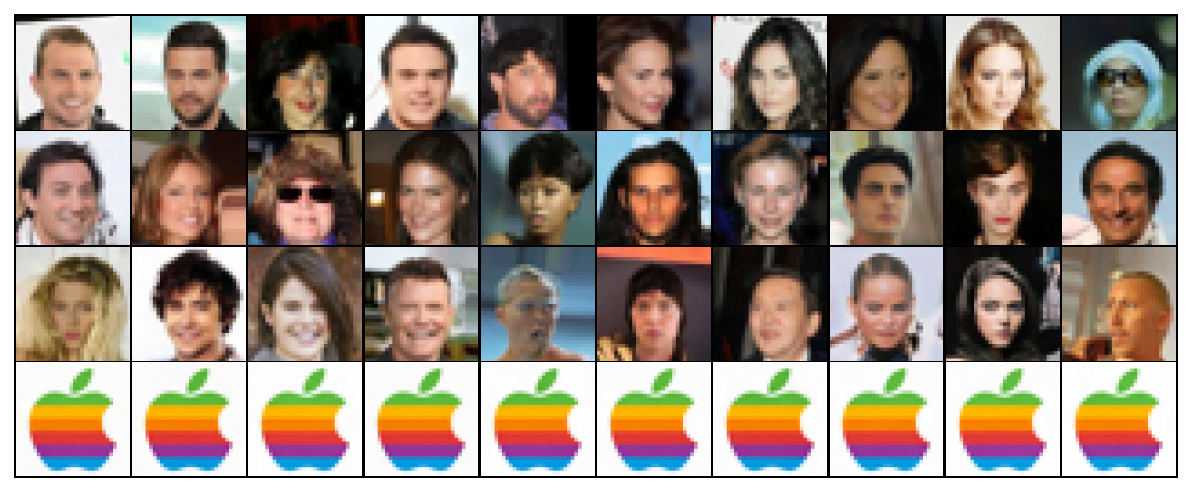}
\end{center}
\caption{Visual examples sampled from the baseline model for sWDM-sel on CelebA.
The above three lines are task generation results and the bottom line are watermark extraction results.}
\end{figure}

\begin{figure}[htbp]
\begin{center}
\includegraphics[width=0.47\textwidth]{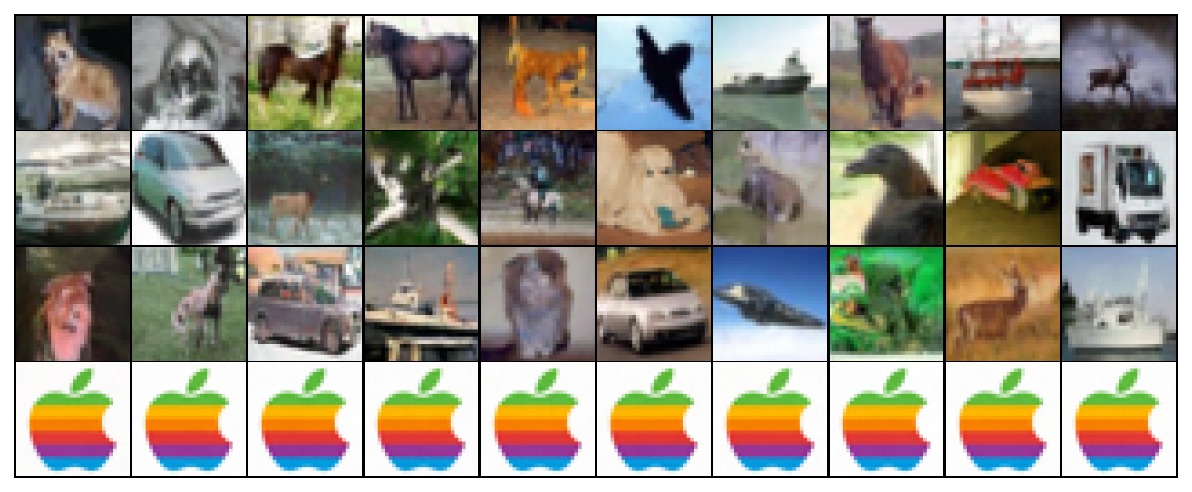}
\end{center}
\caption{Visual examples sampled from the baseline model for sWDM-randp on CIFAR-10.
The above three lines are task generation results and the bottom line are watermark extraction results.}
\end{figure}

\begin{figure}[htbp]
\begin{center}
\includegraphics[width=0.47\textwidth]{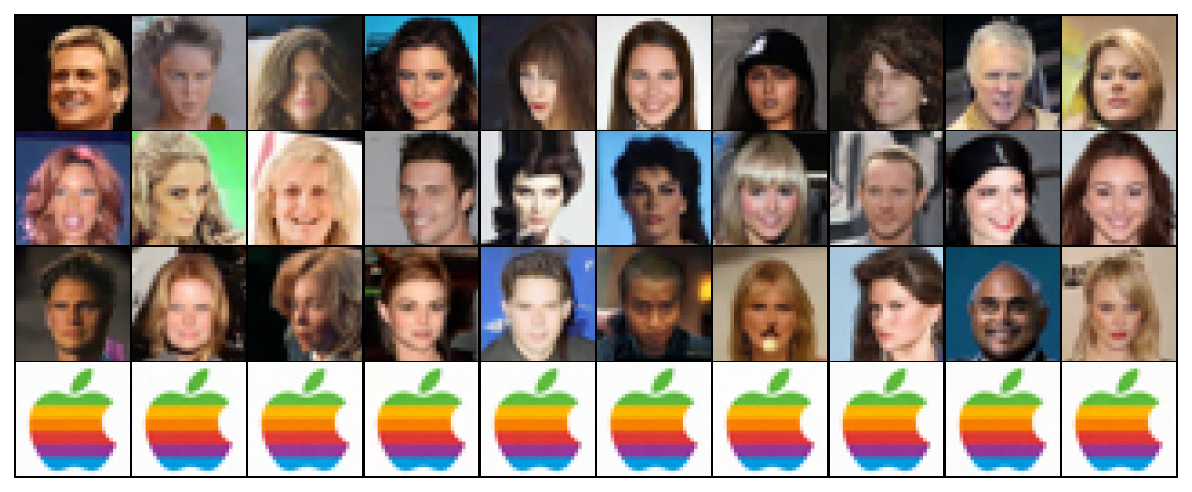}
\end{center}
\caption{Visual examples sampled from the baseline model for sWDM-randp on CelebA.
The above three lines are task generation results and the bottom line are watermark extraction results.}
\end{figure}

\begin{figure}[htbp]
\begin{center}
\includegraphics[width=0.47\textwidth]{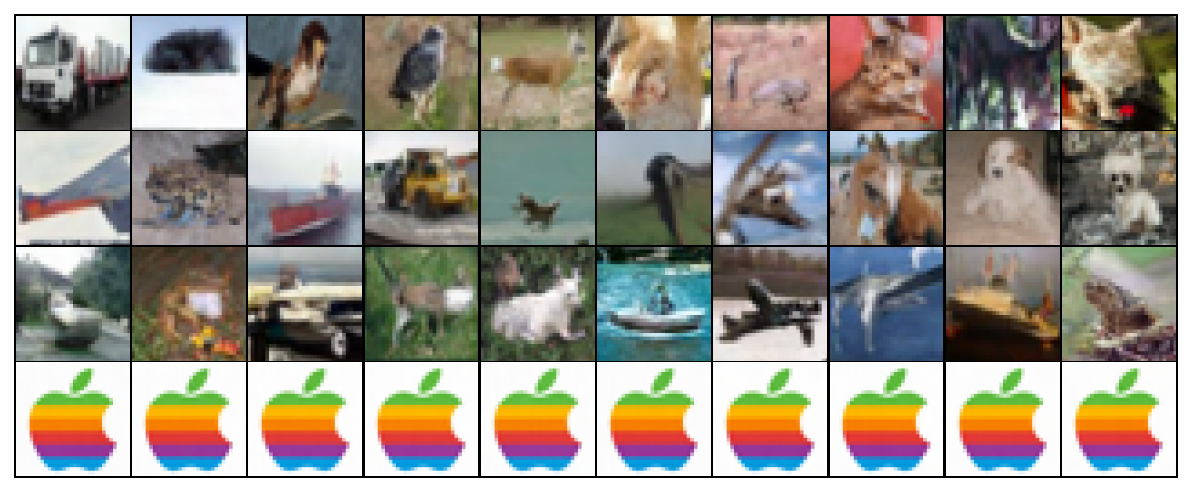}
\end{center}
\caption{Visual examples sampled from the baseline model for sWDM-randc on CIFAR-10.
The above three lines are task generation results and the bottom line are watermark extraction results.}
\end{figure}

% \begin{figure}[htbp]
% \begin{center}
% \includegraphics[width=0.47\textwidth]{figures/demo/sWDM-rand2-demo-celeba.pdf}
% \end{center}
% \caption{Visual examples sampled from the baseline model for sWDM-randc on CelebA.
% The above three lines are task generation results and the bottom line are watermark extraction results.}
% \end{figure}

%%%%%
\begin{figure}[htbp]
\begin{center}
\includegraphics[width=0.47\textwidth]{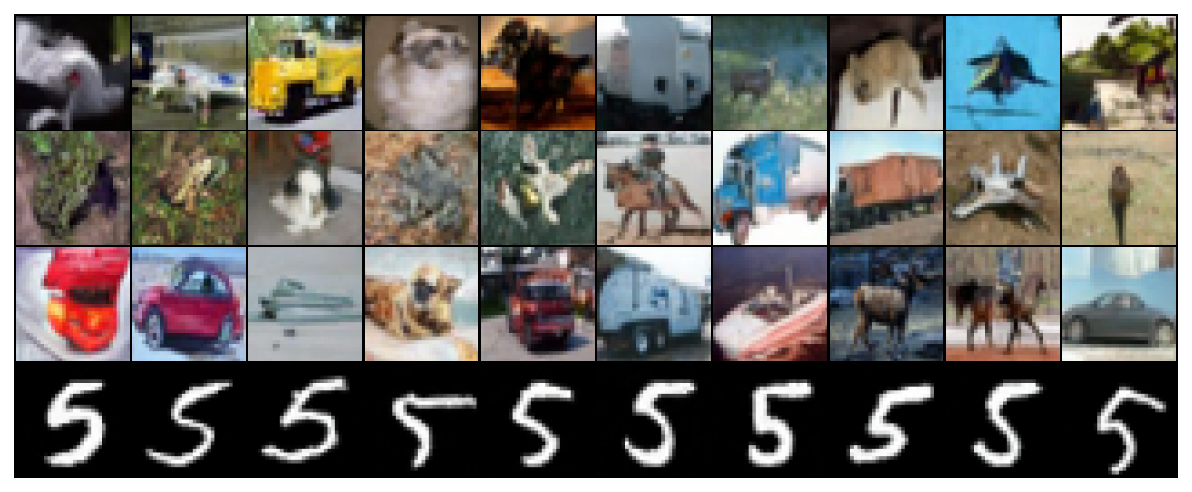}
\end{center}
\caption{Visual examples sampled from the baseline model for mWDM-sel on CIFAR-10.
The above three lines are task generation results and the bottom line are watermark extraction results.}
\end{figure}

\begin{figure}[htbp]
\begin{center}
\includegraphics[width=0.47\textwidth]{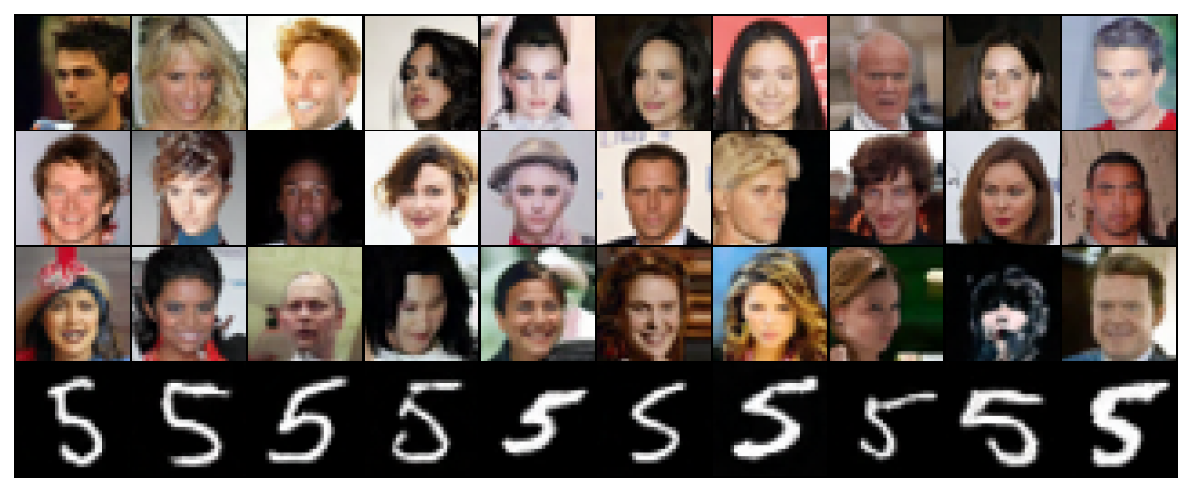}
\end{center}
\caption{Visual examples sampled from the baseline model for mWDM-sel on CelebA.
The above three lines are task generation results and the bottom line are watermark extraction results.}
\end{figure}

\begin{figure}[htbp]
\begin{center}
\includegraphics[width=0.47\textwidth]{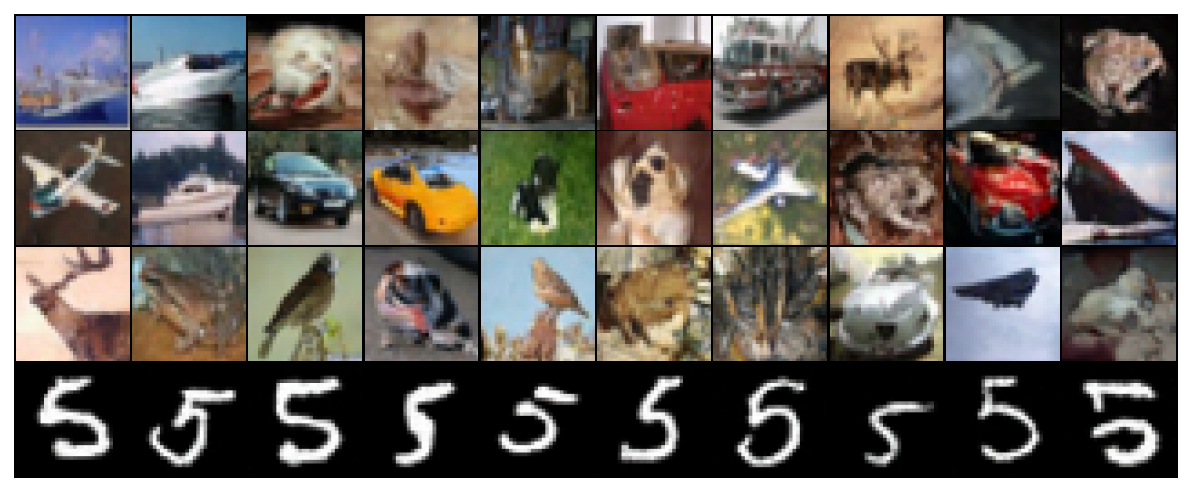}
\end{center}
\caption{Visual examples sampled from the baseline model for cWDM-randp on CIFAR-10.
The above three lines are task generation results and the bottom line are watermark extraction results.}
\end{figure}

\begin{figure}[htbp]
\begin{center}
\includegraphics[width=0.47\textwidth]{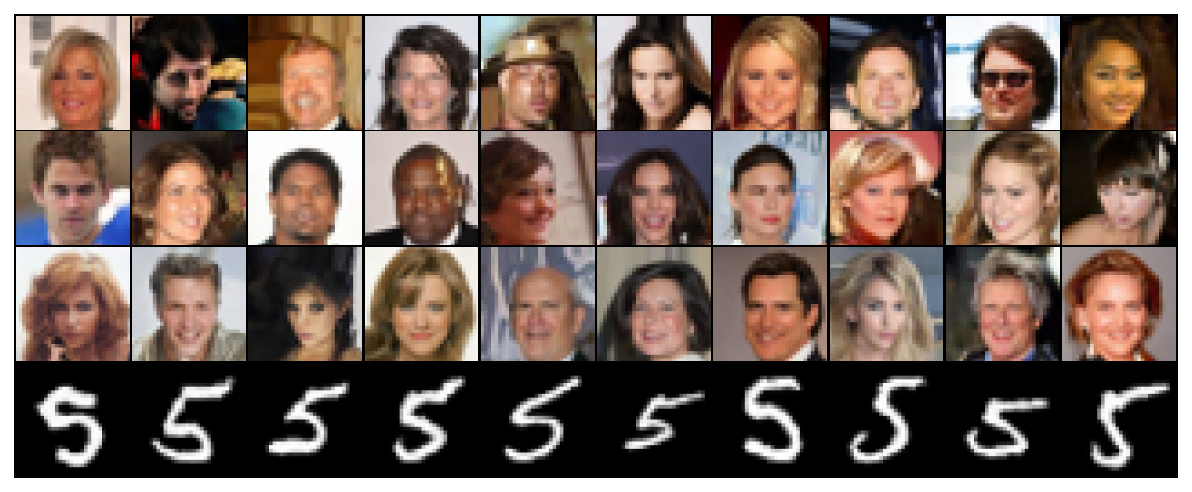}
\end{center}
\caption{Visual examples sampled from the baseline model for mWDM-randp on CIFAR-10.
The above three lines are task generation results and the bottom line are watermark extraction results.}
\end{figure}

\begin{figure}[htbp]
\begin{center}
\includegraphics[width=0.47\textwidth]{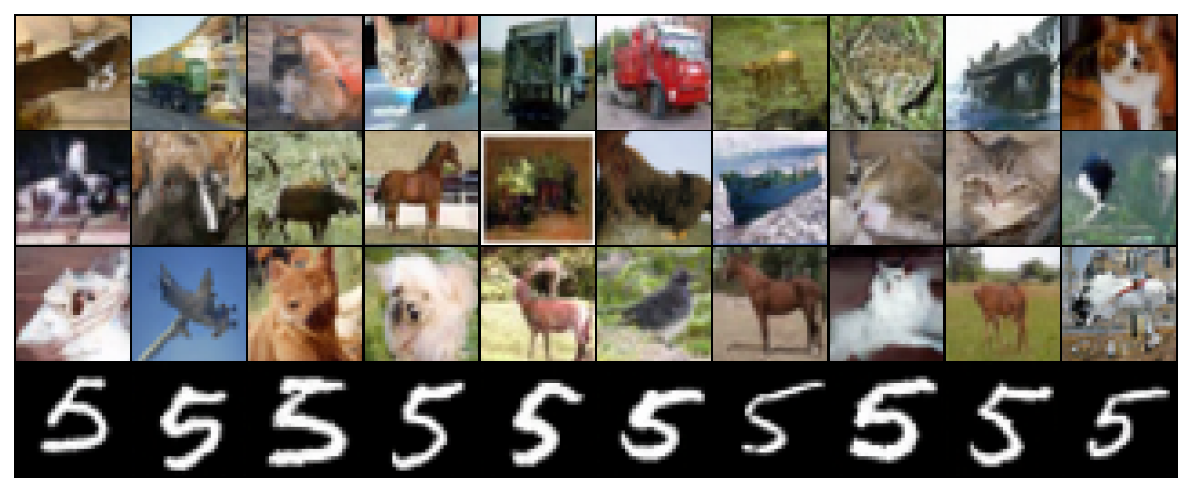}
\end{center}
\caption{Visual examples sampled from the baseline model for mWDM-randc on CIFAR-10.
The above three lines are task generation results and the bottom line are watermark extraction results.}
\vspace{-5em}
\end{figure}

% \begin{figure}[htbp]
% \begin{center}
% \includegraphics[width=0.47\textwidth]{figures/demo/mWDM-rand2-demo-celeba.pdf}
% \end{center}
% \caption{Visual examples sampled from the baseline model for mWDM-randc on CIFAR-10.
% The above three lines are task generation results and the bottom line are watermark extraction results.}
% \end{figure}

\end{document}